\documentclass[a4paper,12pt]{article}
\usepackage{amsmath}
\usepackage{latexsym}

\begin{document}
\begin{titlepage}
\begin{flushright}
gr-qc/0008050\\
\end{flushright}

\begin{centering}
\vspace*{1.cm}

{\large{\bf Time-Dependent Automorphism Inducing Dif\mbox{}feomorphisms}}\\
{\large{\bf In Vacuum Bianchi Cosmologies}}\\
{\large{\bf And The Complete Closed Form Solutions For Type II \&
V}}

\vspace*{1.cm} {\bf T. Christodoulakis\footnote{tchris@cc.uoa.gr},
G. Kofinas\footnote{gkofin@phys.uoa.gr}, E. Korfiatis, G.O. Papadopoulos
\footnote{gpapado@cc.uoa.gr}\\
and A. Paschos}

\vspace{.7cm} {\it University of Athens, Physics Department
Nuclear \& Particle Physics Section\\
GR--15771  Athens, Greece}\\
\end{centering}

\vspace{1.cm} \numberwithin{equation}{section} \pagestyle{empty}
\abstract{We investigate the set of spacetime general coordinate
transformations (G.C.T.) which leave the line element of a generic
Bianchi Type Geometry, quasi-form invariant; i.e. preserve
manifest spatial Homogeneity. We find that these G.C.T.'s, induce
special time-dependent automorphic changes, on the spatial scale
factor matrix $\gamma_{\alpha\beta}(t)$ --along with corresponding
changes on the lapse function $N(t)$ and the shift vector
$N^{\alpha}(t)$. These changes, which are Bianchi Type dependent,
form a group and are, in general, different from those induced by
the group SAut(G) advocated in earlier investigations as the
relevant symmetry group; they are used to simplify the form of the
line element --and thus simplify Einstein's equations as well--,
without losing generality. As far as this simplification procedure
is concerned, the transformations found, are proved to be
essentialy unique. For the case of Bianchi Types II and V, where
the most general solutions are known --Taub's and Joseph's,
respectively--, it is explicitly verified that our transformations
and only those, suffice to reduce the generic line element, to the
previously known forms. It becomes thus possible, --for these
Types-- to give in closed form, the most general solution,
containing all the necessary ''gauge'' freedom.}
\end{titlepage}

\newpage

\section{Introduction}
It is well known, that spatial homogeneity reduces Einstein's
field equations for pure gravity, to a system of ten coupled
O.D.E.'s with respect to time \cite{1}: one equation quadratic in
the velocities $\dot{\gamma}_{\alpha\beta}$  and algebraic in
$N^{2} ~(G_{00}=0)$, three linear in velocities and also algebraic
in $N^{\alpha} ~(G_{0i}=0)$, and the six spatial equations
$(G_{ij}=0)$ which are linear in $\ddot{\gamma}_{\alpha\beta}$ and
are also involving $N, \dot{N}, N^{\alpha}, \dot{N}^{\alpha},
\gamma_{\alpha\beta}, \dot{\gamma}_{\alpha\beta}$.\\
In attempting to find solutions to this set of equations, it is
natural --although seldom adopted in the literature-- to solve the
quadratic constraint for $N^{2}$ and the linear constraint
equations for as many of the $N^{\alpha}$'s as possible; then
substitute into the remaining spatial equations. When this is
done, the spatial equations can be solved for only 6-4=2
independent accelerations $\ddot{\gamma}_{\alpha\beta}$. Only for
Bianchi Type II and III --a particular VI case-- we can solve for
6-3=3 accelerations, since only two of the three linear
constraints are independent; but then in both of these cases, a
linear combination of the $N^{a}$'s remains arbitrary and
counterbalances the existence of the third independent
acceleration. Thus, the general solution to the above mentioned
system of equations will, in every Bianchi Type, involve four
arbitrary functions of time, whose specification should, somehow,
correspond to a choice of time and space coordinates --in complete
analogy to the full pure gravity, whereby four arbitrary functions
of
the spacetime coordinates, are expected to enter the general solution.\\
In the literature a different approach is more frequently met. It
involves an a priori gauge choice of coordinate system: As far as
time is concerned, one may set $N$ to be either an explicit
function of time --say 1 or $t^{2}$ e.t.c.--, or some combination
of $\gamma_{\alpha\beta}$'s --see (2.8). For the spatial
coordinates, the depicted situation is more vague. In some works,
$N^{\alpha}$'s are set to zero, in others, some $N^{\alpha}$'s are
retained. In any case, most of these choices, are considered as
being, more or less, inequivalent and their connection to the
well-known existence of Gauss-normal coordinates $(g_{00}=-1,
~g_{0i}=0)$ \cite{2}, is not at all clear. When such a gauge
choice has been made, the spatial equations can be solved for all
6 independent $\ddot{\gamma}_{\alpha\beta}(t)$. The constraint
equations become then algebraic equations, restricting the initial
data --needed to specify a particular solution of the spatial equations.\\
In both approaches, the ensuing equations are still too difficult
to handle; thus further simplifying hypotheses are employed, such
as $N^{\alpha}(t)=0$, leading to
$\gamma_{\alpha\beta}=diag((a^{2}(t)$, $b^{2}(t), c^{2}(t)))$ for
Class A Types e.t.c. For the Bianchi types I and IX, the
hypothesis $N^{\alpha}(t)=0$ and
$\gamma_{\alpha\beta}=diag((a^{2}(t), b^{2}(t), c^{2}(t)))$, is
known to be linked to kinematics and/or dynamics --although in a,
somewhat, vague way see e.g. \cite{3} and Ryan in \cite{1}. In all
other cases, this or any other simplifying hypothesis used, is
interpreted only as an ansatz to be tested at the end, i.e. after
having solved all the --further simplified-- equations. For
example, to take an extreme case, diagonality of
$\gamma_{\alpha\beta}(t)$ together with the vanishing of the shift
vector is known to lead to incompatibility for Bianchi types IV,
VII (Class B) \cite{4,5}, as well as for the biaxial type VIII
cases $(a^{2}, a^{2}, c^{2}), (a^{2}, b^{2}, a^{2})$, \cite{5}.
The diversity of the various ansatzen appearing in the literature,
causes a considerable degree of fragmentation.

It has long been suspected and/or known, that automorphisms, ought
to play an important role in a unified treatment of this problem.
The first mention, goes back to the first of \cite{6}. More
recently, Jantzen, --second of \cite{6}-- has used Time Dependent
Automorphism Matrices, as a convenient parametrization of a
general positive definite $3\times 3$ scale factor matrix
$\gamma_{\alpha\beta}(t)$, in terms of a --desired-- diagonal
matrix. His approach, is based on the orthonormal frame bundle
formalism, and the main conclusion is (third of [6], pp. 1138):
''\textit{\ldots the special automorphism matrix group SAut(G), is
the symmetry group of the ordinary differential equations,
satisfied by the metric matrix $\gamma_{\alpha\beta}$, when no
sources are present \ldots}'' Later on, Samuel and Ashtekar in
\cite{7}, have seen automorphisms, as a result of general
coordinate transformations. Their spacetime point of view, has led
them, to consider the --so called-- ''Homogeneity Preserving
Diffeomorphisms'', and link them, to topological considerations.

In this paper, we also take a spacetime point of view, and try to
avoid the fragmentation --mentioned above--, by revealing those
G.C.T.'s, which enable us to simplify the line-element --and
therefore Einstein's equations--, while at the same time, they
preserve manifest spatial homogeneity. We are, thus, able to
identify special automorphic transformations of
$\gamma_{\alpha\beta}(t)$, along with corresponding --non
tensorial, for the shift vector-- changes of $N, N^{\alpha}$ which
allow us to set $N^{\alpha}=0$ and bring $\gamma_{\alpha\beta}(t)$
to some irreducible, simple --though not unique-- form.

The structure of the paper, is organized as follows:\\
In section 2, after establishing the existence of Time-Dependent
Automorphism Inducing Diffeomorphisms (A.I.D.'s), the general
irreducible form of the line element for all Bianchi Types is
given, and a uniqueness theorem, is proven.\\
In section 3, attention is focused on Bianchi Types II and V.
Einstein's equations obtaining from the irreducible form of the
line element, are explicitly written down and completely
integrated. The uniqueness of the transformations given in
section 2, is explicitly verified, with the aid of the well known
Taub's and Joseph's solution --respectively. As a result, we give
the closed form of the most general line elements,
satisfying equations (2.5).\\
Finally, some concluding remarks are included in the discussion.

\section{Time Dependent Automorphism\\ Inducing Dif\mbox{}feomorphisms}
It is well known that the vacuum Einstein field equations can be
derived from an action principle:
\begin{equation}
\mathcal{A}=\frac{-1}{16\pi}\int\sqrt{-^{(4)}g}~^{(4)}R~d^{4}x
\end{equation}
(we use geometrized units i.e. $G=c=1$)\\
The standard canonical formalism \cite{8} makes use of the lapse
and shift functions appearing in the 4-metric:
\begin{equation}
ds^{2}=(N^{i} N_{i}-
N^{2})dt^{2}+2N_{i}dx^{i}dt+g_{ij}dx^{i}dx^{j}
\end{equation}
From this line-element the following set of equations obtains,
expressed in terms of the extrinsic curvature:
\begin{displaymath}
K_{ij}=\frac{1}{2N}(N_{i\mid j}+N_{i\mid j}-\frac{\partial g_{ij}
}{\partial t})
\end{displaymath}
\begin{subequations}
\begin{equation}
H_{0}=\sqrt{g}~(K_{ij}K^{ij}-K^{2}+R)=0
\end{equation}
\begin{equation}
H_{i}=2\sqrt{g}~(K^{j}_{i\mid j}-K_{\mid i})=0
\end{equation}
\begin{equation}
\begin{split}
\frac{1}{\sqrt{g}}\frac{d}{dt}[\sqrt{g}~(K^{ij}-K g^{ij})]=-N
(R^{ij}-\frac{1}{2}R~
g^{ij})-\frac{N}{2}(K_{kl}K^{kl}-K^{2})g^{ij}\\
+2N(K^{ik}K^{j}_{k}-K~K^{ij})- (N^{\mid ij}-N^{\mid l}_{\mid
l}g^{ij})+[(K^{ij}-K~g^{ij})N^{l}]_{\mid l}\\-N^{i}_{\mid
l}(K^{lj}-K~g^{lj})-N^{j}_{\mid l}(K^{li}-K~g^{li})
\end{split}
\end{equation}
\end{subequations}
This set is equivalent to the ten Einstein's equations.

In cosmology, we are interested in the class of spatially
homogeneous spacetimes, characterized by the existence of an
m-dimensional isometry group of motions $G$, acting transitively
on each surface of simultaneity $\Sigma_{t}$. When m is greater
than 3 and there is no proper invariant subgroup of dimension 3,
the spacetime is of the Kantowski-Sachs type \cite{9} and will not
concern us further. When m equals the dimension of $\Sigma_{t}$
--which is 3--, there exist 3 basis one-forms
$\sigma^{\alpha}_{i}$ satisfying:
\begin{subequations}
\begin{equation}
d\sigma^{\alpha}=C^{\alpha}_{\beta\gamma}~\sigma^{\beta}\wedge
\sigma^{\gamma}\Leftrightarrow \sigma^{\alpha}_{i
,~j}-\sigma^{\alpha}_{j,~i}=2C^{\alpha}_{\beta\gamma}~
\sigma^{\gamma}_{i}~\sigma^{\beta}_{j}
\end{equation}
where $C^{\alpha}_{\beta\gamma}$ are the structure constants of
the corresponding isometry group.\\
In this case there are local coordinates $t, ~x^{i}$ such that the
line element in (2.2) assumes the form:
\begin{equation}
\begin{split}
ds^{2}&=(N^{\alpha}(t) N_{\alpha}(t)-
N^{2}(t))dt^{2}+2N_{\alpha}(t)\sigma^{\alpha}_{i}(x)dx^{i}dt\\&
+\gamma_{\alpha\beta}\sigma^{\alpha}_{i}(x)\sigma^{\beta}_{j}(x)dx^{i}dx^{j}
\end{split}
\end{equation}
\end{subequations}
Latin indices, are spatial with range from 1 to 3. Greek indices,
number the different basis 1-forms, take values in the same range,
and are lowered and raised by $\gamma_{\alpha\beta}$, and
$\gamma^{\alpha\beta}$ respectively.

A commitment concerting the topology of the $3$-surface, is
pertinent here, especially in view of the fact that we wish to
consider diffeomorphisms \cite{7}; we thus assume that $G$ is
simply connected and the $3$-surface $\Sigma_{t}$ can be
identified with $G$, by singling out a point $p$ of $\Sigma_{t}$,
as the identity $e$, of $G$.

If we insert relations (2.4) into equations (2.3), we get the
following set of ordinary differential equations for the
Bianchi-Type spatially homogeneous spacetimes:
\begin{subequations}
\begin{equation}
E_{0}\doteq K^{\alpha}_{\beta}~K^{\beta}_{\alpha}-K^{2}+R=0
\end{equation}
\begin{equation}
E_{\alpha}\doteq
K^{\mu}_{\alpha}~C^{\epsilon}_{\mu\epsilon}-K^{\mu}_{\epsilon}~C^{\epsilon}_{\alpha\mu}=0
\end{equation}
\begin{equation}
E^{\alpha}_{\beta}\doteq
\dot{K}^{\alpha}_{\beta}-NKK^{\alpha}_{\beta}+NR^{\alpha}_{\beta}+
2N^{\rho}(K^{\alpha}_{\nu}~C^{\nu}_{\beta\rho}-K^{\nu}_{\beta}~C^{\alpha}_{\nu\rho})
\end{equation}
\end{subequations}
where $K^{\alpha}_{\beta}=\gamma^{\alpha\rho}K_{\rho\beta}$ and
\begin{equation}
K_{\alpha\beta}=-\frac{1}{2N}(\dot{\gamma}_{\alpha\beta}+2\gamma_{\alpha\nu}C^{\nu}_{\beta\rho}N^{\rho}+2\gamma_{\beta\nu}C^{\nu}_{\alpha\rho}N^{\rho})
\end{equation}
\begin{equation}
\begin{split}
R_{\alpha\beta}&=C^{\kappa}_{\sigma\tau}C^{\lambda}_{\mu\nu}\gamma_{\alpha\kappa}\gamma_{\beta\lambda}\gamma^{\sigma\nu}\gamma^{\tau\mu}+
2C^{\lambda}_{\alpha\kappa}C^{\kappa}_{\beta\lambda}+2C^{\mu}_{\alpha\kappa}C^{\nu}_{\beta\lambda}\gamma_{\mu\nu}\gamma^{\kappa\lambda}\\
&+2C^{\lambda}_{\beta\kappa}C^{\mu}_{\mu\nu}\gamma_{\alpha\lambda}\gamma^{\kappa\nu}+
2C^{\lambda}_{\alpha\kappa}C^{\mu}_{\mu\nu}\gamma_{\beta\lambda}\gamma^{\kappa\nu}
\end{split}
\end{equation}
When $N^{\alpha}=0$, equation (2.5c) reduces to the form of the
equation given in \cite{10}. Equation set (2.5), forms what is
known as a --complete-- perfect ideal; that is, there are no
integrability conditions obtained from this system. So, with the
help of (2.5c), (2.6), (2.7), it can explicitly be shown, that the
time derivatives of (2.5a) and (2.5b) vanish identically. The
calculation is staightforward --although somewhat lengthy. It
makes use of the Jacobi identity
$C^{\alpha}_{\rho\beta}C^{\rho}_{\gamma\delta}+C^{\alpha}_{\rho\delta}C^{\rho}_{\beta\gamma}+
C^{\alpha}_{\rho\gamma}C^{\rho}_{\delta\beta}=0$, and its
contracted form
$C^{\alpha}_{\alpha\beta}C^{\beta}_{\gamma\delta}=0$.

The vanishing of the derivatives of the 4 constrained equations:\\
$E_{0}=0, E_{\alpha}=0$, implies that these equations, are first
integrals of equations (2.5c) --moreover, with vanishing
integration constants. Indeed, algebraically solving (2.5a),
(2.5b) for $N(t), N^{\alpha}(t)$, respectively and substituting in
(2.5c), one finds that in all --but Type II and III-- Bianchi
Types, equations (2.5c), can be solved for only 2 of the 6
accelerations $\ddot{\gamma}_{\alpha\beta}$ present. In Type II
and III, the independent accelerations are 3, since $E_{\alpha}$
are not independent and thus, can be solved for only 2 of  the 3
$N^{\alpha}$'s. But then in both of these cases, a linear
combination of the $N^{a}$'s remains arbitrary, and
counterbalances the extra independent acceleration. Thus, in all
Bianchi Types, 4 arbitrary functions of time enter the general
solution to the set of equations (2.5). Based on the intuition
gained from the full theory, one could expect this fact to be a
reflection of the only known covariance of the theory; i.e. of the
freedom to make arbitrary changes of the time and space
coordinates.

The rest of this section is devoted to the investigation of the
existence, uniqueness, and properties of general coordinate
transformations --containing 4 arbitrary functions of time--,
which on the one hand, must preserve the manifest spatial
homogeneity, of the line element (2.4b), and on the other hand,
must be symmetries of equations (2.5).\\
As far as time reparametrization is concerned the situation is
pretty clear: If a transformation
\begin{subequations}
\begin{equation}
t\rightarrow \tilde{t}=g(t) \Leftrightarrow t=f(\tilde{t })
\end{equation}
is inserted in the line element (2.4b), it is easily inferred that
\begin{equation}
\gamma_{\alpha\beta}(t)\rightarrow
\gamma_{\alpha\beta}(f(\tilde{t}))\equiv
\tilde{\gamma}_{\alpha\beta}(\tilde{t})
\end{equation}
\begin{equation}
\begin{split}
N(t)\rightarrow \pm~
N(f(\tilde{t}))\frac{df(\tilde{t})}{d\tilde{t}}\equiv
\widetilde{N}(\tilde{t})\\
N^{\alpha}(t)\rightarrow
N^{\alpha}(f(\tilde{t}))\frac{df(\tilde{t})}{d\tilde{t}}\equiv
\widetilde{N}^{\alpha}(\tilde{t})
\end{split}
\end{equation}
\end{subequations}
Accordingly, $K^{\alpha}_{\beta}$ transforms under (2.8a) as a
scalar and thus (2.5a), (2.5b) are also scalar equations while
(2.5c) gets multiplied by a factor $df(\tilde{t})/d\tilde{t}$.
Thus, given a particular solution to equations (2.5), one can
always obtain an equivalent solution, by arbitrarily redefining
time. Hence, we understand the existence of one arbitrary function
of time in the general solution to Einstein's equations (2.5). In
order to understand the presence of the rest 3 arbitrary functions
of time it is natural to turn our attention to the tranformations
of the 3 spatial coordinates $x^{i}$. To begin with, consider the
transformation:
\begin{equation}
\begin{split}
\tilde{t}=t\Leftrightarrow& t=\tilde{t}\\
\tilde{x}^{i}=g^{i}(x^{j},t)\Leftrightarrow&
x^{i}=f^{i}(\tilde{x}^{j},\tilde{t})
\end{split}
\end{equation}
It is here understood, that our previous assumption concerning the
topology of $G$ and the identification of $\Sigma_{t}$ with $G$,
is valid for all values of the parameter $t$, for which the
transformation is to be well defined.

Under these transformations, the line element (2.4b) becomes:
\begin{equation}
\begin{split}
ds^{2}&=[(N^{\alpha}N_{\alpha}-N^{2})+\frac{\partial
f^{i}}{\partial \tilde{t}}\frac{\partial f^{j}}{\partial \tilde{t}
}~\sigma^{\alpha}_{i}(f)\sigma^{\beta}_{j}(f)\gamma_{\alpha\beta}(\tilde{t})\\
&+2\sigma^{\alpha}_{i}(f)\frac{\partial f^{i}}{\partial
\tilde{t}}N_{\alpha}(\tilde{t})]d\tilde{t}^{2}\\
&+2\sigma^{\alpha}_{i}(x)\frac{\partial x^{i}}{\partial
\tilde{x}^{m}}[N_{\alpha}(\tilde{t})+\sigma^{\beta}_{j}(x)\frac{\partial
x^{j}}{\partial
\tilde{t}}\gamma_{\alpha\beta}(\tilde{t})]d\tilde{x}^{m}d\tilde{t}\\
&+\sigma^{\alpha}_{i}(x)\sigma^{\beta}_{j}(x)\gamma_{\alpha\beta}(\tilde{t})\frac{\partial
x^{i}}{\partial \tilde{x}^{m}}\frac{\partial x^{j}}{\partial
\tilde{x}^{n}}~d\tilde{x}^{m}d\tilde{x}^{n}
\end{split}
\end{equation}
Since our aim, is to retain manifest spatial homogeneity of the
line element (2.4b), we have to refer the form of the line element
given in (2.10) to the old basis $\sigma^{\alpha}_{i}(\tilde{x})$
at the new spatial point $\tilde{x}^{i}$. Since
$\sigma^{\alpha}_{i}$ --both at $x^{i}$ and $\tilde{x}^{i}$--, as
well as, $\partial x^{i}/
\partial \tilde{x}^{j}$, are invertible matrices, there
always exists a non-singular matrix $\Lambda
^{\alpha}_{\mu}(\tilde{x},\tilde{t})$ and a triplet
$P^{\alpha}(\tilde{x},\tilde{t})$, such that:
\begin{equation}
\begin{split}
\sigma^{\alpha}_{i}(x)\frac{\partial x^{i}}{\partial
\tilde{x}^{m}}=&\Lambda^{\alpha}_{\mu}(\tilde{x},\tilde{t})\sigma^{\mu}_{m}(\tilde{x})\\
\sigma^{\alpha}_{i}(x)\frac{\partial x^{i}}{\partial
\tilde{t}}=&P^{\alpha}(\tilde{x},\tilde{t})
\end{split}
\end{equation}
The above relations, must be regarded as definitions, for the
matrix $\Lambda ^{\alpha}_{\mu}$ and the triplet $P^{\alpha}$.
With these identifications the line element (2.10) assumes the
form:
\begin{equation}
\begin{split}
ds^{2}&=[(N^{\alpha}N_{\alpha}-
N^{2})+P^{\alpha}(\tilde{x},\tilde{t})P^{\beta}(\tilde{x},\tilde{t})\gamma_{\alpha\beta}(\tilde{t})
+2P^{\alpha}(\tilde{x},\tilde{t})N_{\alpha}(\tilde{t})]d\tilde{t}^{2}\\
&+2\Lambda^{\alpha}_{\mu}(\tilde{x},\tilde{t})\sigma^{\mu}_{m}(\tilde{x})[N_{\alpha}(\tilde{t})
+P^{\beta}(\tilde{x},\tilde{t})\gamma_{\alpha\beta}(\tilde{t})]d\tilde{x}^{m}d\tilde{t}\\
&+\Lambda^{\alpha}_{\mu}(\tilde{x},\tilde{t})\Lambda^{\beta}_{\nu}(\tilde{x},\tilde{t})
\gamma_{\alpha\beta}(\tilde{t})\sigma^{\mu}_{m}(\tilde{x})\sigma^{\nu}_{n}(\tilde{x})d\tilde{x}^{m}
d\tilde{x}^{n}
\end{split}
\end{equation}
If, following the spirit of \cite{7}, we wish the transformation
(2.9) to be manifest homogeneity preserving i.e. to have a well
defined, non-trivial action on $\gamma_{\alpha\beta}(t)$, $N(t)$
and $N^{\alpha}(t)$, we must impose the condition that
$\Lambda^{\alpha}_{\mu}(\tilde{x},\tilde{t})$ and
$P^{\alpha}(\tilde{x},\tilde{t})$ do not depend on the spatial
point $\tilde{x}$, i.e.
$\Lambda^{\alpha}_{\mu}=\Lambda^{\alpha}_{\mu}(\tilde{t})$ and
$P^{\alpha}=P^{\alpha}(\tilde{t})$. Then (2.12) is written as:
\begin{equation}
\begin{split}
ds^{2}&=[(N^{\alpha}N_{\alpha}-
N^{2})+P^{\alpha}P^{\beta}\gamma_{\alpha\beta}
+2P^{\alpha}N_{\alpha}]d\tilde{t}^{2}\\
&+2\Lambda^{\alpha}_{\mu}\sigma^{\mu}_{m}(\tilde{x})[N_{\alpha}
+P^{\beta}\gamma_{\alpha\beta}]d\tilde{x}^{m}d\tilde{t}\\
&+\Lambda^{\alpha}_{\mu}\Lambda^{\beta}_{\nu}
\gamma_{\alpha\beta}\sigma^{\mu}_{m}(\tilde{x})\sigma^{\nu}_{n}(\tilde{x})d\tilde{x}^{m}
d\tilde{x}^{n}\Rightarrow\\
ds^{2}&\equiv (\widetilde{N}^{\alpha}\widetilde{N}_{\alpha}-
\widetilde{N}^{2})d\tilde{t}^{2}+2\widetilde{N}_{\alpha}(\tilde{t})\sigma^{\alpha}_{i}(\tilde{x})d\tilde{x}^{i}d\tilde{t}\\
&+\widetilde{\gamma}_{\alpha\beta}(\tilde{t})\sigma^{\alpha}_{i}(\tilde{x})\sigma^{\beta}_{j}(\tilde{x})d\tilde{x}^{i}d\tilde{x}^{j}
\end{split}
\end{equation}
with the allocations:
\begin{subequations}
\begin{equation}
\widetilde{\gamma}_{\alpha\beta}=\Lambda^{\mu}_{\alpha}\Lambda^{\nu}_{\beta}\gamma_{\mu\nu}
\end{equation}
\begin{equation}
\widetilde{N}_{\alpha}=\Lambda^{\beta}_{\alpha}(N_{\beta}+P^{\rho}\gamma_{\rho\beta})~~
and~~thus~~\widetilde{N}^{\alpha}=S^{\alpha}_{\beta}(N^{\beta}+P^{\beta})
\end{equation}
\begin{equation}
\widetilde{N}=N
\end{equation}
\end{subequations}
(where $S=\Lambda^{-1}$).\\
Of course, the demand that $\Lambda^{\alpha}_{\beta}$ and
$P^{\alpha}$ must not depend on the spatial point $\tilde{x}^{i}$,
changes the character of (2.11), from identities, to the following
set of differential restrictions on the functions defining the
transformation:
\begin{subequations}
\begin{equation}
\frac{\partial f^{i}}{\partial
\tilde{x}^{m}}=\sigma^{i}_{\alpha}(f)\Lambda^{\alpha}_{\beta}(\tilde{t})\sigma^{\beta}_{m}(\tilde{x})
\end{equation}
\begin{equation}
\frac{\partial f^{i}}{\partial
\tilde{t}}=\sigma^{i}_{\alpha}(f)P^{\alpha}(\tilde{t})
\end{equation}
\end{subequations}
Equations (2.15) constitute a set of first-order highly non-linear
P.D.E.'s for the unknown functions $f^{i}$. The existence of local
solutions to these equations is guaranteed by Frobenius theorem
\cite{11} as long as the necessary and sufficient conditions:
\begin{displaymath}
\frac{\partial}{\partial \tilde{x}^{j}}\Big( \frac{\partial
f^{i}}{\partial \tilde{x}^{m}}\Big)- \frac{\partial}{\partial
\tilde{x}^{m}}\Big( \frac{\partial f^{i}}{\partial
\tilde{x}^{j}}\Big)=0
\end{displaymath}
\begin{displaymath}
\frac{\partial}{\partial \tilde{t}}\Big( \frac{\partial
f^{i}}{\partial \tilde{x}^{m}}\Big)- \frac{\partial}{\partial
\tilde{x}^{m}}\Big( \frac{\partial f^{i}}{\partial
\tilde{t}}\Big)=0
\end{displaymath}
hold. Through (2.15) and repeated use of (2.4a), these equations
reduce respectively to:
\begin{equation}
\Lambda^{\alpha}_{\mu}C^{\mu}_{\beta\gamma}=\Lambda^{\rho}_{\beta}\Lambda^{\sigma}_{\gamma}C^{\alpha}_{\rho\sigma}
\end{equation}
\begin{equation}
P^{\mu}C^{\alpha}_{\mu\nu}\Lambda^{\nu}_{\beta}=\frac{1}{2}\dot{\Lambda}^{\alpha}_{\beta}
\end{equation}
It is noteworthy that the solutions to (2.16) and (2.17), --by virtue of (2.14)--
form a group, with composition law:
\begin{displaymath}
(\Lambda_{3})^{\alpha}_{\beta}=(\Lambda_{1})^{\alpha}_{\varrho}(\Lambda_{2})^{\varrho}_{\beta}
\end{displaymath}
\begin{displaymath}
(P_{3})^{a}=(\Lambda_{1})^{\alpha}_{\beta}(P_{2})^{\beta}+(P_{1})^{a}
\end{displaymath}
where ($\Lambda_{1},~P_{1}$) and ($\Lambda_{2},~P_{2}$), are two
successive transformations of the form (2.14).\\
Note also, that a constant automorphism is always a solution of
(2.16), (2.17); indeed,
$\Lambda^{a}_{\beta}(t)=\Lambda^{a}_{\beta}$ and $P^{a}(t)=0$
solve these equations. Thus, $\Lambda^{a}_{\beta}$ and $P^{a}=0$
can be regarded as the remaining gauge symmetry, after one has
fully used the arbitrary functions of time, appearing in a
solution $\Lambda^{a}_{\beta}(t)$ and $P^{a}(t)$. Consequently one
can, at first sight, regard all the arbitrary constants
encountered when integrating (2.17), as absorbable in the shift,
since the transformation law for the shift, is then tensorial.
This is certainly true, as long as there is a non zero initial
shift. However, if one has used the independent functions of time,
in order to set the shift zero, then the constants remaining
within $\Lambda^{a}_{\beta}$, are not absorbable. It is this kind
of constants that we explicitly present below, when we give the
solutions to (2.16), (2.17) for all Bianchi Types. A relevant nice
discussion, distinguishing between genuine gauge symmetries (cf.
arbitrary functions of time) and rigid symmetries (cf. arbitrary
constants), is presented in \cite{12}. There a different
definition of manifest homogeneity preserving diffeomorphisms
--stronger than the one adopted in this work-- is used, and
results in only the inner automorphisms being allowed to acquire
$t$ dependence. In connection to this, it is interesting to
observe that (2.16-17) give essentially the same results: notice
that $2P^{\mu}C^{\alpha}_{\mu\beta}$ is, by definition, the
generator of Inner Automorphisms. Thus there is always a
$\lambda^{\alpha}_{\beta}(t)\equiv
Exp(2P^{\mu}C^{\alpha}_{\mu\beta})$ $\in$ IAut(G) satisfying
(2.17). If we now parameterize the general solution to (2.16-17)
by
$\Lambda^{\alpha}_{\beta}(t)=\lambda^{\alpha}_{\varrho}(t)U^{\varrho}_{\beta}(t)$
and substitute in these relations, we deduce that the matrix $U$
is a constant automorphism. This analysis is verified in the
explicit solutions to (2.16-17), presented latter.

Equation (2.16) is satisfied if and only if,
$\Lambda^{\alpha}_{\beta}(t)$ is an element of the automorphism
group of the Lie algebra determined by the
$C^{\alpha}_{\beta\gamma}$. Equation (2.17) further restricts the
form of $\Lambda^{\alpha}_{\beta}(t)$ and $P^{\alpha}(t)$, so that
manifest spatial homogeneity is preserved despite the mixing of
the old time and space coordinates in the new spatial coordinates
$\tilde{x}^{i}$. Thus, it is appropriate to call transformations
(2.9), satisfying conditions (2.15), (2.16), (2.17),
\textbf{Time-Dependent Automorphism Inducing Diffeomorphisms}. The
importance of automorphisms in Bianchi Cosmologies, has been
stressed in \cite{6}. The symmetry group of the differential
equations, satisfied by $\gamma_{\alpha\beta}(t)$, --advocated in
these works of Jantzen \textit{et al}-- is the unimodular matrices
SAut(G). As we shall later see, we find another symmetry group,
whose time-dependent part lies essentially in IAut(G) and thus
coincides with SAut(G), only for
Class A Bianchi Types VI, VII, VIII, IX.\\
At this point it is natural to ask how this difference occurs. It
is our opinion that the key elements on which the difference in
the symmetry groups found rests, are:
\begin{itemize}
\item[a)] The inhomogeneous transformation law (2.14b) for the
shift. Indeed, Jantzen (1979), having adopted an orthonormal
frame-bundle point of view, naturally assumes as his ''gauge''
transformation laws (2.14a,c) and the tensorial law\\
$\bar{N}^{a}=S^{a}_{\beta}N^{\beta}$, for the shift p. 221.
\item[b)] The different definition and/or role reserved for the
triplet $P^{a}(t)$; we define it as a sort of ''velocity'' of the
transformation (2.9) in (2.15b) and use it in the inhomogeneous
law (2.14b). On the other hand, Jantzen (1979), (p. 221) uses the
corresponding quantity $\omega^{a}(t)$ (so called velocity of the
automorphism frame)to define a new time derivative
$\partial/\partial \bar{t}=\partial/\partial
t+\omega^{a}(t)\sigma^{i}_{a}(x)\partial/\partial x^{i}$.
\item[c)] We concentrate on the symmetries of the O.D.E.s (2.5),
i.e. of \emph{Eistein's equations written in the invariant base},
while Jantzen, as far as we understand, focuses on the symmetries
of the P.D.E.s (2.3), i.e. \emph{of Einstein's equations, written
in an arbitrary frame}.
\end{itemize}

In \cite{7}, the so called Homogeneity Preserving Diffeomorphisms,
are considered in relation to the topology of $\Sigma_{t}$. A
time-independent version of (2.15), appears in \cite{13}, where
all homogeneous three-geometries, are classified in equivalence
classes, with respect to these ''frozen'' transformations. It is
straightforward to check, that $E_{0}, ~ E_{\alpha},~
E^{\alpha}_{\beta}$ transform --under (2.14)-- as follows:
\begin{equation}
\widetilde{E}_{0}=E_{0},
~~\widetilde{E}_{\alpha}=\Lambda^{\beta}_{\alpha}E_{\beta},
~~\widetilde{E}^{\alpha}_{\beta}=S^{\alpha}_{\mu}\Lambda^{\nu}_{\beta}E^{\mu}_{\nu}
\end{equation}
This fact, establishes the covariance of equations (2.5), under
the ''gauge'' transformation (2.14), and implies that if
$(N,~N^{\alpha},~\gamma_{\alpha\beta})$ is a solution to
Einstein's equations, so will be the set
$(\widetilde{N},~\widetilde{N}^{\alpha},~\widetilde{\gamma}_{\alpha\beta})$
--provided that, (2.16), (2.17) hold--; in fact, as the preceding
exposition proves, they will be the same equations expressed in
different space-time coordinate systems. Out of the twelve
quantities $\Lambda^{\alpha}_{\beta}(t)$ and $P^{\alpha}(t)$,
conditions (2.16), (2.17) leave us, as we are going to see, in
every Bianchi Type, with 3 arbitrary functions of time. This fact,
along with the time reparametrization covariance, completes our
understanding of why four arbitrary functions of time enter the
general solution to (2.5). Consequently, transformation (2.14),
gives us the possibility to simplify the form of the line element,
and thus of Einstein's equations without loss of generality. It is
obvious, that the simplification obtained, is different for
different Bianchi Types, and even within a particular Bianchi Type
it is not unique --since one may ''spend'' the freedom of the
three arbitrary functions in different ways.

A particularly interesting result, is that the shift vector
$\widetilde{N}^{\alpha}$ can always be put to zero --perhaps at
the expense of a more complicated
$\widetilde{\gamma}_{\alpha\beta}$. For the sake of completeness,
we give below, a detailed analysis of the space of solutions to
(2.16) and (2.17), for each and every Bianchi Type (solutions
to (2.16), have been presented in \cite{14}). \\
To this end, recall that in 3 dimensions, the tensor
$C^{\alpha}_{\beta\gamma}$, admits a unique decomposition in terms
of a contravariant symmetric tensor density of weight -1,
$m^{\alpha\beta}$ and a covariant vector
$\nu_{\alpha}=\frac{1}{2}C^{\rho}_{\alpha\rho}$ as follows
\cite{15}:
\begin{displaymath}
C^{\alpha}_{\beta\gamma}=m^{\alpha\delta}\varepsilon_{\delta\beta\gamma}+\nu_{\beta}\delta^{\alpha}_{\gamma}-\nu_{\gamma}\delta^{\alpha}_{\beta}
\end{displaymath}
The contracted Jacobi identities imply that
$m^{\alpha\beta}\nu_{\beta}=0$, i.e. $\nu_{\alpha}$ is a null
eigenvector of the matrix $m^{\alpha\beta}$. Under a $GL(3,\Re)$
linear mixing of the basis 1-forms
$\sigma^{\alpha}\rightarrow\widetilde{\sigma}^{\alpha}=S^{\alpha}_{\beta}\sigma^{\beta}$,
the structure constant tensor transforms as:
\begin{displaymath}
C^{\alpha}_{\beta\gamma}\rightarrow
\widetilde{C}^{\alpha}_{\beta\gamma}=S^{\alpha}_{\mu}\Lambda^{\nu}_{\beta}\Lambda^{\rho}_{\gamma}C^{\mu}_{\nu\rho}
\end{displaymath}
Accordingly, the $m^{\alpha\beta}$ and $\nu_{\alpha}$ transform
as:
\begin{displaymath}
\widetilde{m}^{\alpha\beta}=|S|^{-1}S^{\alpha}_{\gamma}S^{\beta}_{\delta}m^{\gamma\delta}
\end{displaymath}
\begin{displaymath}
\widetilde{\nu}_{\alpha}=\Lambda^{\beta}_{\alpha}\nu_{\beta}
\end{displaymath}
$\Lambda$ (and thus $S$) is called a Lie algebra automorphism if
$C^{\alpha}_{\beta\gamma}=\widetilde{C}^{\alpha}_{\beta\gamma}$,
i.e. if $\widetilde{m}^{\alpha\beta}$ and
$\widetilde{\nu}_{\alpha}$ are equal to $m^{\alpha\beta}$ and
$\nu_{\alpha}$ respectively. In this case the automorphism
conditions become:
\begin{subequations}
\begin{equation}
m^{\alpha\beta}=|S|^{-1}S^{\alpha}_{\gamma}S^{\beta}_{\delta}m^{\gamma\delta}
\end{equation}
\begin{equation}
\nu_{\alpha}=\Lambda^{\beta}_{\alpha}\nu_{\beta}
\end{equation}
\end{subequations}

The different Bianchi Types, arise according to the rank and
signature of $m^{\alpha\beta}$, and the vanishing or not, of
$\nu_{\alpha}$. Using (2.19), one can --straightforwardly-- solve
the system of equations (2.16) and (2.17). We now present, the
form of $\Lambda^{\alpha}_{\beta}(t)$ and $P^{\alpha}(t)$
satisfying (2.16), (2.17) as well as some irreducible form for
$\gamma_{\alpha\beta}$, for each Bianchi Type:

\textbf{Type I}: $m^{\alpha\beta}=0, \nu_{\alpha}=0$. This Type
has been exhaustively treated, in the literature
(\cite{3},~\cite{7}). We only note that --since all
$C^{\alpha}_{\beta\gamma}$ are zero-- (2.17), implies that
$P^{\alpha}(t)$ is arbitrary and $\Lambda^{\alpha}_{\beta}(t)$ is
constant. Then, (2.16) implies that $\Lambda^{\alpha}_{\beta}$ is
an element of $GL(3,\Re)$. Thus, without loss of generality, one
can set $N^{\alpha}=0$, --using (2.14b). A first integral of
equations (2.5c) is then,
$\gamma^{\alpha\rho}\dot{\gamma}_{\rho\beta}=\vartheta^{\alpha}_{\beta}$
where $\vartheta^{\alpha}_{\beta}$, is an arbitrary constant
matrix. From this point, the standard textbooks, \cite{3} deduce
(using algebraic arguments) a diagonal form:\\
$\gamma_{\alpha\beta}=diag(e^{\alpha t}, e^{\beta t}, e^{\gamma
t} )$ and then using Einstein's equations
find the general solution, which depends on 1 essential parameter, as expected --see table.\\
Indeed, from (2.5c), one has 12 initial constants; 6
$\gamma_{\alpha\beta}$, and 6 $\dot{\gamma}_{\alpha\beta}$ at
some $t_{0}$ --according to Peano's theorem. The quadratic
constraint equation (2.5a), reduces them to 10, and then, with the
usage of constant automorphisms --which contain 9
$\Lambda^{\alpha}_{\beta}$'s--, the number of the remaining
essential constants (or essential parameters), is $10-9=1$.

\textbf{Type II} rank($m$)=$1$ and $\nu_{\alpha}=0$ . Then, matrix
$m^{\alpha\beta}$, can be cast to the form
$m^{\alpha\beta}=diag(1/2,0,0)$. Equations (2.16), (2.17) give the
following form for $\Lambda^{\alpha}_{\beta}(t)$:
\begin{displaymath}
\Lambda^{\alpha}_{\beta}(t)=\left( \begin{array}{ccc}
  \varrho_{1}\varrho_{4}-\varrho_{2}\varrho_{3} & x(t) & y(t) \\
  0 & \varrho_{1} & \varrho_{2} \\
  0 & \varrho_{3} & \varrho_{4}
\end{array} \right), ~~~(\varrho_{1},\varrho_{2},\varrho_{3},\varrho_{4}~~
\text{constants})
\end{displaymath}
The triplet $P^{\alpha}(t)$ assumes the form:
\begin{displaymath}
P^{\alpha}(t)=(P(t),~\frac{\varrho_{1}\dot{y}-\varrho_{2}\dot{x}}{\varrho_{1}\varrho_{4}-\varrho_{2}\varrho_{3}},
~\frac{\varrho_{3}\dot{y}-\varrho_{4}\dot{x}}{\varrho_{1}\varrho_{4}-\varrho_{2}\varrho_{3}})
\end{displaymath}
The general solution to this Type, is Taub's solution (\cite{16}),
which contains
2 essential parameters --see table.\\
Again, we can understand this number, using Peano's theorem and
the arbitrary extra constants, appearing in
$\Lambda^{\alpha}_{\beta}$. Using $x(t)$ and $y(t)$, we start with
4 $\gamma_{\alpha\beta}$'s (i.e. we set
$\gamma_{12}=\gamma_{13}=0$) and no shift. Thus the initial
arbitrary constants, are $2\times 4=8$. Out of these, the
quadratic constraint equation (2.5a), removes 2, and 4 more are
eliminated by the 4 $\varrho$'s, contained in
$\Lambda^{\alpha}_{\beta}$. So, the remaining arbitrary constants
are: $8-2-4=2$, in accordance with the number of expected
essential parameters.

\textbf{Type V} rank($m$)=$0$ and $\nu_{\alpha}\neq 0$. Then
$m^{\alpha\beta}=0$ and we may choose
$\nu_{\alpha}=-\frac{1}{2}\delta^{3}_{\alpha}$. Equations (2.16),
(2.17) give the following form for $\Lambda^{\alpha}_{\beta}(t)$:
\begin{displaymath}
\Lambda^{\alpha}_{\beta}(t)=\left( \begin{array}{ccc}
  \varrho_{1}P(t) & \varrho_{2}P(t) & x(t) \\
  \varrho_{3}P(t) & \varrho_{4}P(t) & y(t) \\
  0 & 0 & 1
\end{array} \right), ~~~(\varrho_{1},\varrho_{2},\varrho_{3},\varrho_{4}~~
\text{constants})
\end{displaymath}
with $\varrho_{1}\varrho_{4}-\varrho_{2}\varrho_{3}=1$ and the
triplet:
\begin{displaymath}
P^{\alpha}(t)=(x(\ln \frac{x}{P})^{\cdot},~y(\ln
\frac{y}{P})^{\cdot},~(\ln \frac{1}{P})^{\cdot})
\end{displaymath}
The general solution, is also known, as Joseph's solution
(\cite{17}), with one essential parameter.\\
This number comes naturally, within our method; using $x(t)$ and
$y(t)$, one can eliminate $\gamma_{13}$ and $\gamma_{23}$. Then,
$P(t)$ can serve to set the subdeterminant of
$\gamma_{\alpha\beta}$, equal to $(\gamma_{33})^{2}$. At this
stage, we are left with 3 $\gamma_{\alpha\beta}$'s while the
linear constraints equations (2.5b), imply that the shift is zero.
Again, the quadratic constraint (2.5a), subtracts 2 arbitrary
constants, and the constants contained in
$\Lambda^{\alpha}_{\beta}$, 3 more. Then, the result is:
$6-2-3=1$, essential constant.

\textbf{Type IV} rank($m$)=$1$ and $\nu_{\alpha}\neq 0$. We may
choose\\
$m^{\alpha\beta}=diag(1/2,0,0)$, $
\nu_{\alpha}=-\frac{1}{2}\delta^{3}_{\alpha}$. Equations (2.16),
(2.17) give the following form for $\Lambda^{\alpha}_{\beta}(t)$:
\begin{displaymath}
\Lambda^{\alpha}_{\beta}(t)=\left( \begin{array}{ccc}
  P(t) & P(t)\ln[\kappa P(t)] & x(t) \\
  0 & P(t) & y(t) \\
  0 & 0 & 1
\end{array} \right), ~~~(\kappa~~
\text{constant})
\end{displaymath}
and the triplet:
\begin{displaymath}
P^{\alpha}(t)=(x(\ln \frac{x}{P})^{\cdot}-\dot{y},~y(\ln
\frac{y}{P})^{\cdot},~(\ln \frac{1}{P})^{\cdot})
\end{displaymath}
In this Type --which is a class B Type--, we can set
$\gamma_{13}=\gamma_{23}=0$, using $x(t)$ and $y(t)$. At this
stage, the 2 of the 3 linear constraint equations, imply
$N^{1}=N^{2}=0$, while the third, involves $P(t)$. Thus we can
further, either set $N^{3}=0$ --through (2.14b)-- and retain a
non-zero $\gamma_{12}$, or eliminate $\gamma_{12}$, at the expense
of a non-vanishing $N^{3}$. It is well known, that
$N^{3}=0$ and $\gamma_{12}=0$, leads to incompatibility [4].\\
We have thus, the following counting of the essential parameters:\\
a) When $\gamma_{12} \neq 0$ and $N^{3}=0$, we have 8 $-~2$ (from
the quadratic constraint) $-~2$ (from the remaining linear
equation) $-~1$ (from the constant contained in
$\Lambda^{\alpha}_{\beta}$) $=3$.\\
b) When $\gamma_{12} = 0$ and $N^{3} \neq 0$, we have 6 $-~2$
(from the quadratic constraint) $-~1$ (from the constant contained
in $\Lambda^{\alpha}_{\beta}$) $=3$. Notice that here, the
remaining linear constraint equation, simply serves to define
$N^{3}$ and thus, does not remove any constant.

\textbf{Type VI} (Including Type III \cite{19},~\cite{18})
~~rank($m$)=$2$, signature($m$)=Lorentzian and $\nu_{\alpha}\neq
0$. One convenient choice is $m^{\alpha\beta}=diag(1,-1,0)$ and
$\nu_{\alpha}=h\delta^{3}_{\alpha}$.\\
\textit{Note: In the standard texts e.g. \cite{15}, the matrix
$m^{a\beta}$ is given in a more complicated form, which carries
part of the arbitrariness of the magnitude of the vector
$\nu_{\alpha}$. In this work, we imply that
$C^{\alpha}_{\beta\gamma}$ are given by their defining relation in
terms of $~\varepsilon_{\alpha\beta\gamma}, ~m^{\alpha\beta},
~\nu_{\alpha}$.}

For all values of $h\neq 0,\pm 1$, equations (2.16), (2.17) give
the following form for $\Lambda^{\alpha}_{\beta}(t)$:
\begin{displaymath}
\Lambda^{\alpha}_{\beta}(t)=\left( \begin{array}{ccc}
  e^{-hP(t)}\lambda\cosh(P(t)) & e^{-hP(t)}\lambda\sinh(P(t)) & x(t) \\
  e^{-hP(t)}\lambda\sinh(P(t)) & e^{-hP(t)}\lambda\cosh(P(t)) & y(t) \\
  0 & 0 & 1
\end{array} \right)
\end{displaymath}
\begin{displaymath}
~~~(\lambda~~ \text{constant})
\end{displaymath}
while the triplet:
\begin{displaymath}
\begin{split}
P^{\alpha}(t)=&(-\frac{(h^{2}-1)x(t)\dot{P}(t)+h\dot{x}(t)+\dot{y}(t)}{2(h^{2}-1)},\\
&-\frac{(h^{2}-1)y(t)\dot{P}(t)+h\dot{y}(t)+\dot{x}(t)}{2(h^{2}-1)},-\frac{\dot{P}(t)}{2})
\end{split}
\end{displaymath}

For $h=0$, --class A--, there are two solutions:
\begin{displaymath}
\Lambda^{\alpha}_{\beta}(t)=\left( \begin{array}{ccc}
  \lambda\cosh(P(t)) & \lambda\sinh(P(t)) & x(t) \\
  \epsilon\lambda\sinh(P(t)) & \epsilon\lambda\cosh(P(t)) & y(t) \\
  0 & 0 & \epsilon
\end{array} \right)
\end{displaymath}
\begin{displaymath}
~~~(\lambda~~ \text{constant})
\end{displaymath}
while the triplet:
\begin{displaymath}
P^{\alpha}(t)=(\frac{\epsilon\dot{y}(t)-x(t)\dot{P}(t)}{2},\frac{\epsilon\dot{x}(t)-y(t)\dot{P}(t)}{2},-\frac{\epsilon\dot{P}(t)}{2})
\end{displaymath}
where $\epsilon=\pm 1$.

For $h=\pm 1$, --class B--, the solutions are:
\begin{displaymath}
\Lambda^{\alpha}_{\beta}(t)=\left( \begin{array}{ccc}
  e^{-hP(t)}\lambda\cosh(P(t)) & e^{-hP(t)}\lambda\sinh(P(t)) & x(t) \\
  e^{-hP(t)}\lambda\sinh(P(t)) & e^{-hP(t)}\lambda\cosh(P(t)) & c-hx(t) \\
  0 & 0 & 1
\end{array} \right)
\end{displaymath}
\begin{displaymath}
~~~(\lambda~~ \text{constant})
\end{displaymath}
while the triplet:
\begin{displaymath}
\begin{split}
P^{\alpha}(t)=&(\Omega(t),\frac{2h\Omega(t)-c\dot{P}(t)+2hx(t)\dot{P}(t)+\dot{x}(t)}{2},\\
&h\frac{e^{-hP(t)}\lambda\sinh(P(t))\dot{P}(t)-he^{-hP(t)}\lambda\cosh(P(t))\dot{P}(t)}{2e^{-hP(t)}\lambda\cosh(P(t))-2he^{-hP(t)}\lambda\sinh(P(t))})
\end{split}
\end{displaymath}
For each of the previously mentioned cases, we have:\\
a) When $h=0$, (class A), $\gamma_{\alpha\beta}$ can be made
diagonal and then the shift vanishes. Thus the counting of the
essential parameters is: 6 $~-2$ (from the quadratic constraint)
$~-1$ (from the constant, contained in
$\Lambda^{\alpha}_{\beta}$) $=3$.\\
b) When $h=\pm 1$, (class B), using $x(t)$ and $P(t)$,  we can
eliminate $\gamma_{13}$ and $\gamma_{23}$. So: 8 $~-2$ (from the
quadratic constraint) $~-2$ (from the constants, contained in
$\Lambda^{\alpha}_{\beta}$) $=4$ is the number of the essential
constants. Notice that the 3 linear constraint equations, are
linearly dependent and thus, when $N^{3}=0$ through (2.14b), there
is no linear constraint equation left, to remove any
constants, hence the number 4.\\
c) When $h\neq 0,\pm1$, the counting algorithm is exactly the
same, as in Type IV case.

\textbf{Type VII} rank($m$)=$2$, signature($m$)=Euclidean and
$\nu_{\alpha}\neq 0$. We may set $m^{\alpha\beta}=diag(-1,-1,0),
~\nu_{\alpha}=h\delta^{3}_{\alpha}$. For all values of $h$,
equations (2.16), (2.17) give the following form for
$\Lambda^{\alpha}_{\beta}(t)$:
\begin{displaymath}
\Lambda^{\alpha}_{\beta}(t)=\left( \begin{array}{ccc}
  \lambda e^{hP(t)}\cos(P(t)) & \lambda e^{hP(t)}\sin(P(t)) & x(t) \\
  -\lambda e^{hP(t)}\sin(P(t)) & \lambda e^{hP(t)}\cos(P(t)) & y(t) \\
  0 & 0 & 1
\end{array} \right)
\end{displaymath}
\begin{displaymath}
~~~(\lambda~~ \text{constant})
\end{displaymath}
and the triplet:
\begin{displaymath}
\begin{split}
P^{\alpha}(t)=&(\frac{x(t)\dot{P}(t)+h^{2}x(t)\dot{P}(t)-h\dot{x}(t)+\dot{y}(t)}{2(1+h^{2})},\\
&\frac{y(t)\dot{P}(t)+h^{2}y(t)\dot{P}(t)-h\dot{y}(t)-\dot{x}(t)}{2(1+h^{2})},\frac{\dot{P}(t)}{2})
\end{split}
\end{displaymath}

For the case $h=0$, there is another solution, except the one
deduced from the previous, by setting $h=0$:
\begin{displaymath}
\Lambda^{\alpha}_{\beta}(t)=\left( \begin{array}{ccc}
  \lambda\cos(P(t)) & \lambda\sin(P(t)) & x(t) \\
  \lambda\sin(P(t)) & -\lambda\cos(P(t)) & y(t) \\
  0 & 0 & -1
\end{array} \right)
\end{displaymath}
\begin{displaymath}
~~~(\lambda~~ \text{constant})
\end{displaymath}
and the triplet:
\begin{displaymath}
P^{\alpha}(t)=(\frac{x(t)\dot{P}(t)-\dot{y}(t)}{2},\frac{y(t)\dot{P}(t)+\dot{x}(t)}{2},-\frac{\dot{P}(t)}{2})
\end{displaymath}

Again, for each of the previously mentioned cases, we have:\\
a) When $h=0$, (class A), $\gamma_{\alpha\beta}$ can be made
diagonal and equations (2.5b) give $N^{a}=0$. Thus: 6 $~-2$ (from
the quadratic constraint) $~-1$ (from the constant, contained in
$\Lambda^{\alpha}_{\beta}$) $=3$ is the number of the
essential constants.\\
b) When $h\neq 0$, the counting algorithm is exactly the same, as
in Type IV case.

\emph{For Bianchi Types VIII and IX, condition (2.17), does not
impose any restriction on $\Lambda^{\alpha}_{\beta}(t)$, but
rather fixes completely, the triplet $P^{a}(t)$, to be:
\begin{displaymath}
P^{a}=\frac{1}{4|m|}\varepsilon_{\beta\tau\kappa}m^{\alpha\beta}\Lambda^{\tau}_{\gamma}\dot{\Lambda}^{\kappa}_{\delta}m^{\gamma\delta}
\end{displaymath}}

\textbf{Type VIII} rank($m$)=$3$, signature($m$)=Lorentzian. A
standard choice is
$m^{\alpha\beta}=\eta^{\alpha\beta}=diag(-1,1,1)$. Since
$|m^{\alpha\beta}|=-1$, (2.19a) implies that
$|\Lambda^{\alpha}_{\beta}|=1$ and thus,
$\Lambda^{\alpha}_{\beta}$'s are the isometries of the Minkowski
metric, in three dimensions, i.e. the Lorentz boosts, with one
timelike and two spacelike directions, times a rotation of the
''space'' plane. Thus, the automorphisms are characterized by the
two components of the velocity vector, plus the rotation angle.
The triplet $P^{a}$, is:
\begin{displaymath}
P^{a}=\frac{1}{2}(\Lambda^{2}_{\mu}\dot{\Lambda}^{3}_{\nu}\eta^{\mu\nu},-\Lambda^{3}_{\mu}\dot{\Lambda}^{1}_{\nu}\eta^{\mu\nu},-\Lambda^{1}_{\mu}\dot{\Lambda}^{2}_{\nu}\eta^{\mu\nu})
\end{displaymath}
It can be proven --see appendix A-- that a positive definite
matrix,
can be diagonalized by this automorphism group; i.e. we can set\\
$\gamma_{\alpha\beta}=diag(a^{2}(t),b^{2}(t),c^{2}(t))$. Then,
from (2.5b), we will have $N^{a}=0$.

The number 4, of the expected essential parameters --see table
below--, can be understood as follows: The time-dependent Lorentz
transformation $ \Lambda^{\alpha}_{\beta}$, can diagonalize
$\gamma_{\alpha\beta}$. Thus, the counting: 6 $-~2$ (from the
quadratic constraint) $=4$.

\textbf{Type IX} rank($m$)=$3$, signature($m$)=Euclidean. The
standard choice is $m^{\alpha\beta}=\delta^{\alpha\beta}$. Since
$|m^{\alpha\beta}|=1$,(2.19a) implies that
$|\Lambda^{\alpha}_{\beta}|=1$ and thus,
$\Lambda^{\alpha}_{\beta}$'s are the isometries of the Euclidean
metric, in three dimensions, i.e. the orthogonal matrices, which
are characterized by three parameters; e.g. the Euler angles. The
triplet $P^{a}$, is:
\begin{displaymath}
P^{a}=\frac{1}{2}(\Lambda^{2}_{\mu}\dot{\Lambda}^{3}_{\nu}\delta^{\mu\nu},\Lambda^{3}_{\mu}\dot{\Lambda}^{1}_{\nu}\delta^{\mu\nu},\Lambda^{1}_{\mu}\dot{\Lambda}^{2}_{\nu}\delta^{\mu\nu})
\end{displaymath}
Since a positive definite symmetric matrix can be diagonalized by
an element of --the connected to the identity component of--
$O(3)$, we have that
$\gamma_{\alpha\beta}(t)=diag(a^{2}(t),~b^{2}(t),~c^{2}(t))$
\cite{19}. Then, from (2.5b), as is well-known $N^{\alpha}=0$.

The counting yields --exactly as in Type VIII: 6 $-~2$ (from the
quadratic constraint) $=4$, essential constants.

>From the above analysis of the space of solutions to (2.16) and
(2.17), we observe that in each Bianchi Type, there are 3
arbitrary functions of time --as expected--, for a twofold
reason;\\
Firstly, because we are solving the integrability conditions for
the existence of a time-dependent spatial diffeomorphism
according to
(2.9).\\
Secondly, because as it has been mentioned in the Introduction,
the system of Einstein's equations (2.5), has a gauge freedom of
4 arbitrary functions of time. But one of them, simply
corresponds to time reparametrization, while the remaining 3, are
the ones we found in the above analysis.\\
In the various Bianchi Types, the 3 arbitrary functions, are
distributed differently among the components of
$\Lambda^{a}_{\beta}(t)$ and $P^{a}(t)$. This fact, together with
the different number of arbitrary constants appearing in
$\Lambda^{a}_{\beta}$ for each Type, results in a different number
of essential constants --expected by independent arguments
\cite{19} to appear in the general solutions to Einstein's
equations (2.5) --see Table.

We now conclude section 2, by stating the following (uniqueness)
Theorem:\\
\emph{''In a given --albeit arbitrary-- Bianchi Type, let
$\gamma_{1}$, $\gamma_{2}$, (in matrix notation) be solutions to
Einstein's equations (2.5), then there is a matrix M of the form:
$M=\Lambda_{1}^{-1}\Sigma\Lambda_{2}$ (where $\Lambda_{1}$ and
$\Lambda_{2}$ are solutions to (2.16) and (2.17) and $\Sigma$,
represents the irrelevant symmetries of the solution in its
irreducible form) which connects them as:
$\gamma_{2}=M^{T}\gamma_{1}M$.''}

\emph{Note}: $N$, $N^{a}$, are understood to be given from the
quadratic and linear constraint equations (2.5a,b).

The proof rests on the previously established facts:\\
a) That the solutions to (2.16) and (2.17), suffice to reduce the
generic $\gamma_{\alpha\beta}$, to a form that will contain the
expected necessary number of essential constants, so as to be
regarded as the most general one --for each and every Bianchi
Type.\\
b) That for every given Bianchi Type, the solutions to (2.16) and
(2.17), form a group.\\
Indeed, let $\gamma_{1}$, $\gamma_{2}$ be solutions to (2.5).
Then there are $\Lambda_{1}$, $\Lambda_{2}$ --along with $P_{1}$,
$P_{2}$ respectively, if needed-- solutions to (2.16) and (2.17),
such that:
\begin{displaymath}
\gamma_{1}=\Lambda^{T}_{1}\gamma_{irreducible}\Lambda_{1}
\end{displaymath}
\begin{displaymath}
\gamma_{2}=\Lambda^{T}_{2}\gamma_{irreducible}\Lambda_{2}
\end{displaymath}
where $\gamma_{irreducible}$, stands for the solution in a form
exhibiting, only the essential constants. From the first of these:
\begin{displaymath}
\gamma_{irreducible}=(\Lambda^{-1}_{1})^{T}\gamma_{1}\Lambda_{1}
\end{displaymath}
Since, --by definition-- $\gamma_{irreducible}$ is a symmetric
matrix there are always, non-trivial matrices $\Sigma$, such that:
\begin{displaymath}
\gamma_{irreducible}=\Sigma^{T}\gamma_{irreducible}\Sigma
\end{displaymath}
Substituting the two last in the second, we obtain:
\begin{displaymath}
\gamma_{2}=(\Lambda_{1}^{-1}\Sigma\Lambda_{2})^{T}\gamma_{1}\Lambda_{1}^{-1}\Sigma\Lambda_{2}
\end{displaymath}
q.e.d.

\section{The Space of Solutions for Type II and V Cases}
In this section, we adopt the more conventional point of view;
that of ''gauge'' fixing, before solving. As far as time is
concerned, we adopt the ''gauge'' fixing condition
$\widetilde{N}=\sqrt{\widetilde{\gamma}}$, since this simplifies
the form of the equations. For the spatial coordinates, as
explained in the previous section, a choice of reference system,
amounts to a choice of time-dependent automorphism --along with a
choice of $P^{\alpha}(t)$--; thus, we select the form of
$\widetilde{\gamma}_{\alpha\beta}(t)$~ to be such that, the linear
equation would imply $\widetilde{N}^{\alpha}=0$. In this
''gauge'', Einstein's equations (2.5) read:
\begin{subequations}
\begin{equation}
-\widetilde{\gamma}^{\alpha\kappa}\widetilde{\gamma}^{\beta\lambda}\dot{\widetilde{\gamma}}_{\kappa\lambda}\dot{\widetilde{\gamma}}_{\alpha\beta}+
(\frac{\dot{\widetilde{\gamma}}}{\widetilde{\gamma}})^{2}-4\widetilde{\gamma}R=0
\end{equation}
\begin{equation}
C^{\epsilon}_{\alpha\mu}\widetilde{\gamma}^{\mu\rho}\dot{\widetilde{\gamma}}_{\rho\epsilon}-C^{\epsilon}_{\mu\epsilon}\widetilde{\gamma}^{\mu\rho}\dot{\widetilde{\gamma}}_{\rho\alpha}=0
\end{equation}
\begin{equation}
\ddot{\widetilde{\gamma}}_{\alpha\beta}-\widetilde{\gamma}^{\mu\nu}\dot{\widetilde{\gamma}}_{\alpha\mu}
\dot{\widetilde{\gamma}}_{\beta\nu}-2\widetilde{\gamma}R_{\alpha\beta}=0
\end{equation}
\end{subequations}
Note that taking the trace of equations (3.1c), one arrives at:
\begin{equation}
(\frac{\dot{\widetilde{\gamma}}}{\widetilde{\gamma}})^{\cdot}-2\widetilde{\gamma}R=0
\end{equation}
A somewhat useful result deriving from (3.2), is the following:
$\widetilde{\gamma}=ae^{\beta t}$ implies $\widetilde{R}=0$, which
is incompatible for all but I Bianchi Types.

We now present, a realization, of the method developed in the
previous section, for the cases of Type II and V, Bianchi
geometries. At this point, a word of caution is pertinent: it is
evident --from the previously mentioned counting, of the expected
number of essential constants--, that the well known Taub's (Type
II) and Joseph's (Type V) solutions, are the most general for the
respective cases \cite{18}. Thus, we should not expect to find
anything new --in this respect. However, the thorough
investigation of the complete space of solutions, requires the
knowledge of the correct (gauge) symmetry group for Einstein's
equations (2.5). In this respect, we shall directly show, that
transformations (2.14), --as specified by conditions (2.16) and
(2.16), applied to Types II and V--, are essentialy, the only
(gauge) symmetries of these Bianchi geometries.

\emph{Note}: From now on we drop the tildes from the various
quantities for simplicity --except in some cases, where
misunderstanding could occur.

\subsection{Bianchi Type II}
As it can be seen, from the results concerning Type II, we can
consider --without loss of generality--, the time-dependent part
$\gamma_{\alpha\beta}(t)$, of the 3-metric, to have the form:
\begin{displaymath}
\gamma_{\alpha\beta}(t)=\left(\begin{array}{ccc}
  a(t) & 0 & 0 \\
  0 & b(t) & f(t) \\
  0 & f(t) & c(t)
\end{array} \right)
\end{displaymath}
It is interesting to observe that, the freedom in arbitrary
functions of time --contained in $\Lambda^{\alpha}_{\beta}(t)$--,
does not suffice to diagonalize $\gamma_{\alpha\beta}(t)$, i.e.
to set $f(t)=0$, a priori. Yet, we know --see (3.16) and (3.17)
below-- that the diagonal Taub's metric, is the irreducible form
of the most general Type II, solution. The reconciliation of
these two, seemingly conflicting facts, obtains only on mass
shell; after we have completely solved (3.1), with
$\gamma_{\alpha\beta}(t)$ given above, $f(t)$ becomes linearly
dependent upon $b(t)$ and $c(t)$, and we can thus, gauge it away
--utilizing the remaining freedom in arbitrary constants,
contained in $\Lambda^{\alpha}_{\beta}(t)$.

\emph{Note}: From now on, we drop the t-symbol --for time
dependence--, from the various quantities; e.g., $a$ stands for
$a(t)$.

Inserting the form of $\gamma_{\alpha\beta}$ in equations (3.1b),
we find that they vanish identically. We next consider, the
following quantity q, which is scalar under a general linear
mixing $\sigma^{\alpha}\rightarrow
\widetilde{\sigma}^{\alpha}=S^{\alpha}_{\beta}\sigma^{\alpha}$,
with $S^{\alpha}_{\beta}~\in ~GL(3,\Re)$,
\begin{displaymath}
q=C^{\kappa}_{\mu\nu}C^{\lambda}_{\tau\sigma}\gamma_{\kappa\lambda}\gamma^{\mu\tau}\gamma^{\nu\sigma}
=\frac{a^{2}}{2\gamma}=\frac{a}{2(bc-f^{2})}
\end{displaymath}
where $\gamma$, as usual, denotes the determinant of the matrix
$\gamma_{\alpha\beta}$. Then, (2.7)., gives the following
non-zero components for the Ricci tensor $R_{\alpha\beta}$, and
the Ricci scalar, $R$:
\begin{equation}
\begin{array}{c}
  R_{11}=-q\gamma_{11}\\
  R_{rs}=q\gamma_{rs} ~~~r,s = 2,3\\
  R=q \\
\end{array}
\end{equation}
The (1,1) component of (3.1c), is an autonomous equation for the
scale factor $a$:
\begin{equation}
(\frac{\dot{a}}{a})^{\cdot}+a^{2}=0
\end{equation}
with a first integral:
\begin{equation}
(\frac{\dot{a}}{a})^{2}+a^{2}=\omega=constant > 0
\end{equation}
Using (3.2), (3.3) and (3.4), we get the equation for q:
\begin{equation}
(\frac{\dot{q}}{q})^{\cdot}+3a^{2}=0
\end{equation}

To obtain first integrals for (3.1c), let us define the new
variables:
\begin{equation}
\begin{array}{cc}
  \overline{\gamma}_{11}=q^{-1/3}\gamma_{11} & \overline{\gamma}_{rs}=q^{1/3}\gamma_{rs} \\
  \overline{\gamma}^{11}=q^{1/3}\gamma^{11} & \overline{\gamma}^{rs}=q^{-1/3}\gamma_{rs}
\end{array}
\end{equation}
Then:
\begin{equation}
\overline{\gamma}=det(\overline{\gamma}_{\alpha\beta})=q^{1/3}\gamma=\frac{a^{2}}{2}q^{-2/3}
\end{equation}
It is straightforward to see that, with the use of (3.7), and
(3.3), (3.4), (3.6), the spatial Einstein's equations (3.1c),
translate into the following simple, integrable, Kasner-like,
equations, --in terms of $\overline{\gamma}_{\alpha\beta}$:
\begin{equation}
(\overline{\gamma}^{\alpha\rho}\dot{\overline{\gamma}}_{\rho\beta})^{\cdot}=0
\end{equation}
with first integrals:
\begin{equation}
  \overline{\gamma}^{\alpha\rho}\dot{\overline{\gamma}}_{\rho\beta}=
  \vartheta^{\alpha}_{\beta}
\end{equation}
where:
\begin{displaymath}
  \vartheta^{\alpha}_{\beta}=\left(\begin{array}{ccc}
    \theta^{1}_{1} & 0 & 0 \\
    0 & \theta & \varrho\\
    0 & \sigma & \pi \
  \end{array}\right)
\end{displaymath}
Taking the trace of (3.10), we obtain --by means of (3.8):
\begin{equation}
2(\frac{\dot{a}}{a}-\frac{1}{3}\frac{\dot{q}}{q})=\vartheta^{a}_{a}=\theta^{1}_{1}+\vartheta^{s}_{s}
\end{equation}
while the (1,1) component of (3.10), gives:
\begin{equation}
\frac{\dot{a}}{a}-\frac{1}{3}\frac{\dot{q}}{q}=\theta^{1}_{1}
\end{equation}
The last two, imply that
$\theta^{1}_{1}=\vartheta^{s}_{s}=\theta+\pi$, so finally, the
matrix $\vartheta$ becomes:
\begin{equation}
  \vartheta^{\alpha}_{\beta}=\left(\begin{array}{ccc}
    \theta+\pi & 0 & 0 \\
    0 & \theta & \varrho\\
    0 & \sigma & \pi
  \end{array}\right)
\end{equation}
Using the relation $\gamma=a^{2}/(2q)$ --earlier mentioned-- as
well as (3.5) (3.6), (3.7) and (3.10), it is straightforward to
see that the quadratic constraint equation (3.1a), becomes a
relation among constants; that is:
\begin{equation}
\omega=2(\vartheta^{s}_{s})^{2}+|\vartheta^{r}_{s}|
\end{equation}

Integrating (3.5), we get the scale factor a:
\begin{equation}
a(t)\doteq a=\frac{\sqrt{\omega}}{\cosh(\pm\sqrt{\omega}t)}
\end{equation}
>From this relation and (3.12), we conclude that:
\begin{equation}
q^{-1/3}a=a_{0}e^{\vartheta^{s}_{s}t}, ~~~a_{0} > 0
\end{equation}

Now utilizing, in matrix notation, the relation:
$\overline{\gamma}\vartheta=\vartheta^{T}\overline{\gamma}$
--which is the consistency requirement for (3.10)-- and (3.16),
we deduce that classical solutions exist, only for matrices
$\vartheta$, with real eigenvalues --and thus diagonalizable.
Since (2.16), (2.17) admit the solutions
$\Lambda^{\alpha}_{\beta}$=constant, $P^{a}=0$, we can invoke a
constant mixing of $\vartheta$, with a matrix of the form:
\begin{displaymath}
\Lambda=\left(\begin{array}{ccc}
  1 & 0 & 0 \\
  0 & \Lambda^{2}_{2} & \Lambda^{2}_{3} \\
  0 & \Lambda^{3}_{2} & \Lambda^{3}_{3}
\end{array}\right)
\end{displaymath}
and reduce it, to a diagonal form. Then, we are essentially led
to the diagonal Taub's solution:
\begin{equation}
\begin{array}{c}
  \overline{\gamma}_{22}=q^{1/3}b=e^{\theta t} \\
  \overline{\gamma}_{33}=q^{1/3}c=e^{\pi t}
\end{array}
\end{equation}

At this point, it is interesting to observe that, if for some
reason, we had not invoked this diagonalizing $\Lambda$, and
instead proceeded with the general $\vartheta^{r}_{s}$, we would
had arrived, at a reducible form of the solutions with a
non-vanishing $\overline{\gamma}_{23}$. However, this
off-diagonal element, can be made to vanish through the action of
the, previously mentioned, $\Lambda$.

Thus, we have shown that ''gauge'' transformations (2.14) --with
(2.16) and (2.17), holding--, suffice to reduce the most general
line element, for the Type II Bianchi Model, to the known Taub's
metric. According to the theorem stated at the end of section 2,
these transformations are, essentially, unique. We are now going
to explicitly verify it --for the case at hand.

A convenient way to proceed, is to start from Taub's form of the
solution and ask ourselves, what is the form of the most general
time-dependent automorphism $\Lambda^{\alpha}_{\beta}(t)$, which
retains the form invariance of Einstein's equations (2.5)
--written in the invariant basis. Since we know that
$\Lambda^{1}_{2}$ and $\Lambda^{1}_{3}$, can be time-dependent,
we focus on a time-dependent matrix $\Lambda$, of the form:
\begin{equation}
\Lambda^{\alpha}_{\beta}=\left(\begin{array}{ccc}
  \varrho & 0 & 0 \\
  0 & \varrho_{1} & \varrho_{2} \\
  0 & \varrho_{3} & \varrho_{4}
\end{array}\right)
\end{equation}
where: $\varrho=\varrho_{1}\varrho_{4}-\varrho_{2}\varrho_{3}$,
and all $\varrho$'s, are time-dependent.

Consider the transformation, induced by this
$\Lambda^{\alpha}_{\beta}$, on $\gamma^{Taub}_{\alpha\beta}$ --in
matrix notation:
\begin{equation}
\widehat{\gamma}=\Lambda^{T}\gamma^{Taub}\Lambda
\end{equation}
The linear constraint equations (3.1b), still imply
$\widehat{N}^{a}=0$. As far as the time gauge fixing condition
$N=\sqrt{\gamma}$ is concerned, we have:
$\sqrt{\widehat{\gamma}}=|\Lambda|\sqrt{\gamma^{Taub}}$, $|
\Lambda|> 0$, and thus:
\begin{displaymath}
\begin{array}{c}
  N^{Taub}dt^{Taub}=\widehat{N}d\widehat{t}\Rightarrow\\
  d\widehat{t}|\Lambda|\sqrt{\gamma^{Taub}}=\sqrt{\gamma^{Taub}}dt^{Taub}\Rightarrow\\
  d\widehat{t}\varrho^{2}=dt^{Taub} \\
\end{array}
\end{displaymath}
Since we wish for the transformation, to be a symmetry of (2.5),
and we have secured that $\widehat{N}^{a}=0$, and selected
$\widehat{N}=\sqrt{\widehat{\gamma}}$, the equation satisfied by
$\widehat{\gamma}_{\alpha\beta}$, would be exactly (3.1) and
(3.2). Only the dot --defining the time derivative, with respect
to Taub's time--, must be replaced by a prime:
\begin{equation}
'\doteq
\frac{d}{d\widehat{t}}=\varrho^{2}(t^{Taub})\frac{d}{dt^{Taub}}=\varrho^{2}(t^{Taub})\times
^{\cdot}
\end{equation}
Defining the corresponding scale quantities
$\widehat{\overline{\gamma}}_{\alpha\beta}$, --according to (3.7)
and (3.8)-- we must have the analogues of (3.10):
\begin{equation}
\widehat{\overline{\gamma}}^{\alpha\rho}\widehat{\overline{\gamma}}~'_{\rho\beta}=\vartheta^{\alpha}_{\beta}
\end{equation}
Equation (3.2), reads:
\begin{equation}
(\frac{\widehat{\gamma}~'}{\widehat{\gamma}})~'-2\widehat{\gamma}\widehat{R}=0
\end{equation}
It also holds:
\begin{equation}
(\frac{\dot{\gamma}_{Taub}}{\gamma_{Taub}})^{\cdot}-2\gamma_{Taub}
R_{Taub}=0
\end{equation}
Translating (3.22) in the $t^{Taub}$-variable, and subtracting
(3.23), we get:
\begin{displaymath}
2(\varrho^{2})^{\cdot\cdot}+(\varrho^{2})^{\cdot}\frac{\dot{\gamma}_{Taub}}{\gamma_{Taub}}=0
\end{displaymath}
which, with the help of $\gamma_{Taub}=a^{2}_{Taub}/2q$, (3.12)
and $(\theta^{1}_{1})^{Taub}=(\vartheta^{s}_{s})^{Taub}$, becomes:
\begin{equation}
2(\varrho^{2})^{\cdot\cdot}+(\varrho^{2})^{\cdot}(2(\vartheta^{s}_{s})^{Taub}-\frac{1}{3}\frac{\dot{q}}{q})=0
\end{equation}
The (1,1) component of (3.21) is:
\begin{displaymath}
\frac{\widehat{\overline{\gamma}}_{11}~'}{\widehat{\overline{\gamma}}_{11}}=\vartheta^{s}_{s}
\end{displaymath}
where:
\begin{displaymath}
\widehat{\overline{\gamma}}_{11}=q^{-1/3}\widehat{\gamma}_{11}=q^{-1/3}\varrho^{2}\gamma^{Taub}_{11}
\end{displaymath}
and thus, that component reads:
\begin{equation}
(\varrho^{2})^{\cdot}+\varrho^{2}(\vartheta^{s}_{s})^{Taub}=\vartheta^{s}_{s}
\end{equation}
Inserting the derivative of (3.25) into (3.24), we have:
\begin{displaymath}
(\varrho^{2})^{\cdot}\frac{\dot{q}}{q}=0
\end{displaymath}
which in conjunction with (3.6), implies $\varrho^{2}$=constant.
Without loss of generality, we can take $\varrho^{2}$=1.
Henceforth, the time variable $\widehat{t}$, may --and will--
denote Taub's time.\\
It is thus left for us, to investigate
the unimodular matrices:
\begin{displaymath}
\left(\begin{array}{ccc}
  1 & 0 & 0 \\
  0 & \varrho_{1} & \varrho_{2} \\
  0 & \varrho_{3} & \varrho_{4}
\end{array}\right)
\end{displaymath}
with: $1=\varrho_{1}\varrho_{4}-\varrho_{2}\varrho_{3}$, and all
$\varrho$'s, are time-dependent.

It can be proved that a convenient parametrization for this task,
is:
\begin{displaymath}
\Lambda^{\alpha}_{\beta}=\left(\begin{array}{cc}
  1 & \begin{array}{cc}
    0 & 0 \
  \end{array} \\
  \begin{array}{c}
    0 \\
    0 \\
  \end{array} & \Lambda^{r}_{s}
\end{array}\right)
\end{displaymath}
where:
\begin{displaymath}
\Lambda^{r}_{s}=R^{r}_{m}L^{m}_{s}
\end{displaymath}
\begin{displaymath}
L^{m}_{s}=\left(\begin{array}{cc}
    \varphi(t) & \chi(t) \\
    0 & 1/\varphi(t) \
  \end{array}\right)
\end{displaymath}
and $R^{r}_{m}$, are the symmetries of the Taub's metric, i.e.\\
$R^{T}\gamma^{Taub}R=\gamma^{Taub}$ --in matrix notation--:
\begin{displaymath}
R=\left(\begin{array}{cc}
  1 & \begin{array}{cc}
    0 & 0 \
  \end{array} \\
  \begin{array}{c}
    0 \\
    0 \\
  \end{array} & R^{r}_{m}
\end{array}\right)
\end{displaymath}
$R^{r}_{m}$ being:
\begin{displaymath}
R^{r}_{m}=\left(\begin{array}{cc}
  \cos(\tilde{g}(t)) & \sin(\tilde{g}(t))e^{-(\kappa-\mu)t/2} \\
  -\sin(\tilde{g}(t))e^{(\kappa-\mu)t/2} & \cos(\tilde{g}(t))
\end{array}\right)
\end{displaymath}
where $\tilde{g}(t)$, is an unspecified function of time, and
$\kappa,~\mu$, the eigenvalues of $\vartheta^{Taub}$.

The system (3.21), gives the following equations for $\chi(t)$ and
$\varphi(t)$:
\begin{subequations}
\begin{equation}
  2\frac{\dot{\varphi}}{\varphi}+\kappa=\theta+\sigma\frac{\chi}{\varphi}
\end{equation}
\begin{equation}
  (\frac{\chi}{\varphi})^{\cdot}=-\sigma(\frac{\chi}{\varphi})^{2}+(\pi-\theta)\frac{\chi}{\varphi}+\varrho
\end{equation}
\begin{equation}
  e^{(\kappa-\mu)t}(\dot{\chi}\varphi-\chi\dot{\varphi})=\frac{\sigma}{\varphi^{2}}
\end{equation}
\begin{equation}
  -2\frac{\dot{\varphi}}{\varphi}+\mu=\pi-\sigma\frac{\chi}{\varphi}
\end{equation}
\end{subequations}

Equation (3.25), for the choice $\varrho^{2}$=1, gives
$(\vartheta^{s}_{s})^{Taub}=\vartheta^{s}_{s}$, and hence:
\begin{equation}
\pi+\theta=\kappa+\mu
\end{equation}
It also implies,
$\widehat{\overline{\gamma}}_{11}=\overline{\gamma}^{Taub}_{11}$,
or $a(t)=a^{Taub}(t)$, as well as, $\omega=\omega^{Taub}$, or
--through (3.14)--:
\begin{equation}
2(\theta+\pi)^{2}+\pi\theta-\varrho\sigma=2(\kappa+\mu)^{2}+\kappa\mu
\xrightarrow{3.27}\kappa\mu=\theta\pi-\varrho\sigma
\end{equation}
Out of the 4 differential equations (3.26), only the first three,
are independent --in view of (3.27). The solution to this system,
for $\sigma\neq0$, is given by:
\begin{equation}
\frac{\chi}{\varphi}=k_{1}-\frac{\lambda^{3}c^{4}e^{-\lambda
t}}{\sigma(1+\lambda^{2}c^{4}e^{-\lambda t})}
~~~\lambda=\kappa-\mu
\end{equation}
--from the Riccati (3.26b), where
$k_{1}=(\pi-\theta+\lambda)/2\sigma$, is the constant special
solution and:
\begin{equation}
\varphi^{2}=\frac{\sigma}{\lambda^{2}c^{2}}(1+\lambda^{2}c^{4}e^{-\lambda
t})~~~\sigma > 0
\end{equation}
Thus, it is easily seen that, (3.29) and (3.30) make the matrix
$L^{m}_{s}$, to be written in the form
$L^{m}_{s}=\Sigma^{m}_{n}\widetilde{L}^{n}_{s}$, where:
\begin{displaymath}
\Sigma^{m}_{n}=\left(\begin{array}{cc}
  \cos(g(t)) & \sin(g(t))e^{-\lambda t/2} \\
  \sin(g(t))e^{\lambda t/2} & -\cos(g(t))
\end{array}\right)
\end{displaymath}
\begin{displaymath}
\widetilde{L}^{n}_{s}=\left(\begin{array}{cc}
  \varepsilon_{1}\frac{\sqrt{\sigma}}{\lambda c} & k_{1}\varepsilon_{1}\frac{\sqrt{\sigma}}{\lambda c} \\
  \varepsilon_{2}c\sqrt{\sigma} &
  (k_{1}-\frac{\lambda}{\sigma})\varepsilon_{2}c\sqrt{\sigma}
\end{array}\right)
\end{displaymath}
with $(\varepsilon_{1})^{2}=(\varepsilon_{2})^{2}=1$, $(\sigma,
~c) > 0$ and:
\begin{displaymath}
\tan(g(t))=\frac{\varepsilon_{2}c^{2}\lambda}{\varepsilon_{1}}e^{-\lambda
t/2}
\end{displaymath}

There are the special cases $\sigma=0$, or $\lambda=0$, which are
easily seen, to fall into the previous case.

Thus, in all cases, there always exist matrices $\Sigma$ and
$\widetilde{L}$, such that the transformation matrix
$\Lambda^{\alpha}_{\beta}$, can be written as:
\begin{displaymath}
\Lambda^{\alpha}_{\beta}=\left(\begin{array}{cc}
  1 & \begin{array}{cc}
    0 & 0 \
  \end{array} \\
  \begin{array}{c}
    0 \\
    0 \\
  \end{array} & R^{r}_{m}\Sigma^{m}_{n}\widetilde{L}^{n}_{s}
\end{array}\right)
\end{displaymath}

This concludes the verification of the Theorem stated at the end
of the section 2, since indeed, $R$ and $\Sigma$, have trivial
action, on $\gamma^{Taub}_{\alpha\beta}$. It is therefore,
evident that the most general $\gamma_{\alpha\beta}$, $N(t)$ and
$N^{a}(t)$, satisfying equations (2.5), are --in matrix notation:
\begin{displaymath}
\gamma_{most
~general}(t)=\Lambda^{T}(t)\gamma^{Taub}(h(t))\Lambda(t)
\end{displaymath}
\begin{displaymath}
\Lambda=\left(\begin{array}{ccc}
  \varrho_{1}\varrho_{4}-\varrho_{2}\varrho_{3} & x(t) & y(t) \\
  0 & \varrho_{1} & \varrho_{2} \\
  0 & \varrho_{3} & \varrho_{4}
\end{array}\right)
\end{displaymath}
where, the $\varrho$'s are constant, and:
\begin{displaymath}
  N(t)=\sqrt{|\gamma_{most ~general}|}\dot{h}(t)
\end{displaymath}
\begin{displaymath}
  N^{\alpha}(t)=S^{\alpha}_{\beta}(t)P^{\beta}(h(t))\dot{h}(t)
\end{displaymath}
\begin{displaymath}
  P^{\beta}(h(t))=\{P(t), \frac{\varrho_{1}\dot{y}(t)-\varrho_{2}\dot{x}(t)}{(\varrho_{1}\varrho_{4}-\varrho_{2}\varrho_{3})\dot{h}(t)}, \frac{\varrho_{3}\dot{y}(t)-\varrho_{4}\dot{x}(t)}{(\varrho_{1}\varrho_{4}-\varrho_{2}\varrho_{3})\dot{h}(t)}\}
\end{displaymath}
\begin{displaymath}
  S=\Lambda^{-1}(t)
\end{displaymath}
\begin{displaymath}
\gamma^{Taub}(h(t))=\left(\begin{array}{ccc}
  a & 0 & 0 \\
  0 & \frac{e^{(2\kappa+\mu)h(t)}}{a} & 0 \\
  0 & 0 & \frac{e^{(\kappa+2\mu)h(t)}}{a}
\end{array}\right)
\end{displaymath}
\begin{displaymath}
a=\frac{\sqrt{\omega}}{\cosh(\pm\sqrt{\omega} h(t))}
\end{displaymath}
\begin{displaymath}
\omega=2(\kappa+\mu)^{2}+\kappa\mu
\end{displaymath}
where the fourth arbitrary function $h(t)$, accounts for the time
reparametrization covariance, i.e. permits us to depart from the
time gauge fixing $N=\sqrt{\gamma}$.

\subsection{Bianchi Type V}
As it can be seen, from the results of section 2, concerning Type
V, we can consider --with the usage of time-dependent A.I.D.'s--,
the time-dependent part $\gamma_{\alpha\beta}(t)$, of the
3-metric, to be of the form:
\begin{displaymath}
\gamma_{\alpha\beta}(t)=\left(\begin{array}{ccc}
  a(t) & b(t) & 0 \\
  b(t) & c(t) & 0 \\
  0 & 0 & f(t)
\end{array} \right)
\end{displaymath}
with $a(t)c(t)-b^{2}(t)=f^{2}(t)$. Again, as it happens for Type
II, the form of the allowed transformation
$\Lambda^{\alpha}_{\beta}(t)$ is such that, one can not set
$b(t)=0$, a priori. Yet, we know --see (3.39) and (3.40) below--
that the diagonal Joseph's metric, is the irreducible form of the
most general Type V, solution. This puzzle, finds its resolution
only on mass shell; after we have completely solved (3.1) with
$\gamma_{\alpha\beta}(t)$ given above, $b(t)$ becomes linearly
dependent upon $a(t)$ and $c(t)$, and we can thus, put it to zero
--utilizing the remaining freedom in arbitrary constants,
contained in $\Lambda^{\alpha}_{\beta}(t)$.

\emph{Note}: From now on, we drop the t-symbol --for time
dependence--, from the various quantities; e.g., $a$ stands for
$a(t)$.

Inserting the form of $\gamma_{\alpha\beta}$ in equations (3.1b),
we find that they vanish identically. We next define, the scalar
--under a general linear mixing $\sigma^{\alpha}\rightarrow
\widetilde{\sigma}^{\alpha}=S^{\alpha}_{\beta}\sigma^{\alpha}$,
with $S^{\alpha}_{\beta}~\in ~GL(3,\Re)$-- quantity q:
\begin{displaymath}
q=C^{\tau}_{\tau\mu}C^{\sigma}_{\sigma\nu}\gamma^{\mu\nu}=\frac{1}{f}
\end{displaymath}
The condition $ac-b^{2}=f^2$, now reads as: $ac-b^{2}=1/q^{2}$, or
\begin{equation}
\gamma=\frac{1}{q^{3}}
\end{equation}
Then, (2.7), gives:
\begin{equation}
\begin{array}{c}
  R_{\alpha\beta}=2q\gamma_{\alpha\beta}\\
  R=6q \\
\end{array}
\end{equation}
The (3,3) component of (3.1c), gives an autonomous equation for
the scalar quantity $q$:
\begin{equation}
(\frac{\dot{q}}{q})^{\cdot}+\frac{4}{q^{2}}=0
\end{equation}
with a first integral:
\begin{equation}
(\frac{\dot{q}}{q})^{2}-\frac{4}{q^{2}}=\omega=constant
\end{equation}

Defining the scaled quantities:
\begin{equation}
\begin{array}{c}
  \overline{\gamma}_{\alpha\beta}=q\gamma_{\alpha\beta} \\
  \overline{\gamma}^{\alpha\beta}=\frac{1}{q}\gamma^{\alpha\beta} \\
  |\overline{\gamma}|=1
\end{array}
\end{equation}
and using (3.32), (3.33), equations (3.1c), are translated into
the following form:
\begin{equation}
(\overline{\gamma}^{\alpha\rho}\dot{\overline{\gamma}}_{\rho\beta})^{\cdot}=0
\end{equation}
with first integrals:
\begin{equation}
\overline{\gamma}^{\alpha\rho}\dot{\overline{\gamma}}_{\rho\beta}=\vartheta^{\alpha}_{\beta}
\end{equation}
where:
\begin{displaymath}
\vartheta^{\alpha}_{\beta}=\left(\begin{array}{ccc}
  \theta & \varrho & 0 \\
  \sigma & -\theta & 0 \\
  0 & 0 & 0
\end{array}\right)
\end{displaymath}
The form of the matrix $\vartheta$, has been derived, using the
form of $\overline{\gamma}_{\alpha\beta}$ and the property that
$|\overline{\gamma}|=1$. Using (3.31), (3.34) and (3.37), the
quadratic constraint (3.1a), becomes a relation, among constants
--as it was expected--, namely:
\begin{equation}
3\omega=\theta^{2}+\varrho\sigma
\end{equation}
The property $ |\overline{\gamma}|=1$, together with the
consistency requirement --in matrix notation--
$\overline{\gamma}\vartheta=\vartheta^{T}\overline{\gamma}$,
which follows from (3.37), enables us to conclude that classical
solutions, exist only for those values of the parameters,
$\theta$, $\varrho$, $\sigma$, for which $\vartheta$, is
diagonalizable, i.e. when $\theta^{2}+\varrho\sigma > 0$.

Since the matrices $\Lambda^{\alpha}_{\beta}$, of the form:
\begin{displaymath}
\Lambda^{\alpha}_{\beta}=\left(\begin{array}{ccc}
  \varrho_{1} & \varrho_{2} & 0 \\
  \varrho_{3} & \varrho_{4} & 0 \\
  0 & 0 & 1
\end{array}\right)
\end{displaymath}
along with $P^{a}=0$, constitute the remaining gauge freedom, we
can invoke such $\Lambda^{\alpha}_{\beta}$, to diagonalize
$\vartheta^{\alpha}_{\beta}$, and --at the same time-- retain the
shift, zero --see (2.14). Now with a diagonal
$\vartheta^{\alpha}_{\beta}$, equations (3.37), essentially imply
that $\overline{\gamma}_{\alpha\beta}$, is diagonal too.

A further integration of (3.34), yields:
\begin{equation}
\frac{1}{f(t)}=q(t)=\left\{ \begin{array}{ll}
  \frac{2}{\sqrt{\omega}}\sinh(\pm\sqrt{\omega}t) & \omega>0 \\
  \pm 2t & \omega=0
\end{array}\right.
\end{equation}
and thus, we are laid to the well known Joseph's solution
--through complete integration of (3.37), for the diagonal case:
\begin{equation}
\begin{array}{c}
  \overline{\gamma}_{11}=qa=e^{\lambda t} \\
  \overline{\gamma}_{22}=qc=e^{-\lambda t} \\
  3\omega=\lambda^{2}>0
\end{array}
\end{equation}
or the Milnor's solution \cite{18}, when $\omega=0$ --with the
corresponding q.

Once again, it is interesting to observe that if, for some
reason, we do not invoke this diagonalizing
$\Lambda^{\alpha}_{\beta}$ and, instead, proceed with the general
$\vartheta^{\alpha}_{\beta}$, we arrive at a reducible form of
the solution, which contains a non-vanishing
$\overline{\gamma}_{12}$. However, this off-diagonal element, can
be made to vanish through the action of the --previously
mentioned-- constant automorphism.

Thus, we have shown, that the ''gauge'' transformations (2.14),
--with (2.16) and (2.17), holding-- suffice to reduce the most
general line element for the Type V Bianchi Model, to the known
Joseph's metric, as predicted from the theorem, stated at the end
of section 2. As we have done for the Type II case, we are now
going to explicitly verify that these transformations, are
essentially unique. To this end, let us consider the most general
time-dependent automorphism, complementary to the time-dependent
automorphism, described in section 2 --for the Type V, case.
\begin{equation}
\Lambda^{\alpha}_{\beta}=\left(\begin{array}{ccc}
  A(t) & B(t) & 0 \\
  C(t) & F(t) & 0 \\
  0 & 0 & 1
\end{array}\right)
\end{equation}
with $A(t)F(t)-B(t)C(t)=1$. The action of such automorphism on
$\gamma^{Joseph}_{\alpha\beta}$, is --in matrix notation:
\begin{displaymath}
\widehat{\gamma}=\Lambda^{T}\gamma^{Joseph}\Lambda
\end{displaymath}
If we insert $\widehat{\gamma}_{\alpha\beta}$, in the linear
constraint equations (3.1b), we learn that $\widehat{N}^{a}$, are
also zero and, since,
$|\widehat{\gamma}_{\alpha\beta}|=|\Lambda|^{2}|\gamma^{Joseph}_{\alpha\beta}|=|\gamma^{Joseph}_{\alpha\beta}|$,
we conclude that we are in the same temporal, as well as spatial,
gauge. Therefore, $\widehat{\gamma}_{\alpha\beta}$, will also
satisfy equations (3.1c). Since $\Lambda^{\alpha}_{\beta}$, is an
automorphism, it is a symmetry of $q$ and thus, if we define the
scaled quantities:
\begin{displaymath}
\widehat{\overline{\gamma}}_{\alpha\beta}=q\widehat{\gamma}_{\alpha\beta}
\end{displaymath}
they must satisfy, the relations analogous to (3.37):
\begin{equation}
\widehat{\overline{\gamma}}^{\alpha\rho}\dot{\widehat{\overline{\gamma}}}_{\rho\beta}=\vartheta^{\alpha}_{\beta}
\end{equation}
where:
\begin{displaymath}
\vartheta^{\alpha}_{\beta}=\left(\begin{array}{ccc}
  \theta & \varrho & 0 \\
  \sigma & -\theta & 0 \\
  0 & 0 & 0
\end{array}\right)
\end{displaymath}
while, $\overline{\gamma}^{Joseph}_{\alpha\beta}$, satisfies the
relations:
\begin{displaymath}
(\overline{\gamma}^{\alpha\rho})_{Joseph}(\dot{\overline{\gamma}}_{\rho\beta})_{Joseph}=(\vartheta^{\alpha}_{\beta})_{Joseph}
\end{displaymath}
where:
\begin{displaymath}
(\vartheta^{\alpha}_{\beta})_{Joseph}=\left(\begin{array}{ccc}
  \lambda & 0 & 0 \\
  0 & -\lambda & 0 \\
  0 & 0 & 0
\end{array}\right)
\end{displaymath}
By virtue of (3.34), --and since $q$, is invariant--, we get that
$\omega=\omega^{Joseph}$, i.e.
\begin{equation}
\theta^{2}+\varrho\sigma=\lambda^{2}
\end{equation}

In order to proceed with the integration of (3.42), it is
convenient, to parametrize $\Lambda^{\alpha}_{\beta}$ in (3.41),
as follows:
\begin{displaymath}
\Lambda^{\alpha}_{\beta}=\left(\begin{array}{cc}
  \Lambda^{r}_{s} & \begin{array}{c}
    0 \\
    0 \\
  \end{array} \\
  \begin{array}{cc}
    0 & 0 \\
  \end{array} & 1
\end{array}\right)
\end{displaymath}
with $ \Lambda^{r}_{s}=R^{r}_{m}L^{m}_{s}$, where $R^{r}_{m}$ is:
\begin{displaymath}
\left(\begin{array}{cc}
  e^{-\lambda t/2} & 0 \\
  0 & e^{\lambda t/2}
\end{array}\right)\cdot\left(\begin{array}{cc}
  \cos(g(t)) & \sin(g(t)) \\
  -\sin(g(t)) & \cos(g(t))
\end{array}\right)\cdot\left(\begin{array}{cc}
  e^{\lambda t/2} & 0 \\
  0 & e^{-\lambda t/2}
\end{array}\right)
\end{displaymath}
i.e. the symmetries of the Joseph's metric; --in matrix notation\\
$R^{T}\gamma^{Joseph}R=\gamma^{Joseph}$ and $L^{m}_{s}$ is:
\begin{displaymath}\left( \begin{array}{cc}
  \varphi(t) & \tau(t) \\
  0 & 1/\varphi(t)
\end{array}\right)
\end{displaymath}

The system (3.42), gives the following differential equations for
$\varphi(t)$ and $\tau(t)$:
\begin{subequations}
\begin{equation}
2\frac{\dot{\varphi}}{\varphi}+\lambda=\theta+\sigma\frac{\tau}{\varphi}
\end{equation}
\begin{equation}
\dot{\tau}\varphi-\tau\dot{\varphi}=\varrho\varphi^{2}-2\theta\varphi\tau-\sigma\tau^{2}
\end{equation}
\begin{equation}
e^{2\lambda
t}(\dot{\tau}\varphi-\tau\dot{\varphi})=\frac{\sigma}{\varphi^{2}}
\end{equation}
\end{subequations}
The solution to this system, for $\sigma\neq 0$, leads to
incompatibility of the form $\varphi^{2}=-e^{2}$, $e$ a function
of time.

For $\sigma=0$, we get:
\begin{equation}
\begin{array}{c}
  \varphi(t)=c_{1}e^{\frac{\theta-\lambda}{2}t} \\
  \tau(t)=c_{1}\frac{\varrho}{2\theta}e^{\frac{\theta-\lambda}{2}t}
\end{array}
\end{equation}
with $c_{1} > 0$, and --from (3.43), for the case at hand--,
$\theta=\pm \lambda$. The case $\theta=\lambda$, trivially gives,
a constant matrix
\begin{displaymath}
L^{m}_{s}=\left(\begin{array}{cc}
  c_{1} & c_{1}\frac{\varrho}{2\lambda} \\
  0 & 1/c_{1}
\end{array} \right)
\end{displaymath}
while, the case $\theta=-\lambda$, gives
\begin{displaymath}
L^{m}_{s}=\left( \begin{array}{cc}
  0 & e^{-\lambda t} \\
  e^{\lambda t} & 0
\end{array} \right)\cdot\left(\begin{array}{cc}
  0 & 1/c_{1} \\
  c_{1} & -c_{1}\frac{\varrho}{2\lambda}
\end{array} \right)
\end{displaymath}
Since the first matrix in the product, is a symmetry of
$(\overline{\gamma}_{\alpha\beta})_{Joseph}$, we again conclude
that, the non-trivial action of $\Lambda^{\alpha}_{\beta}$, on
$(\overline{\gamma}_{\alpha\beta})_{Joseph}$, is tantamount to the
action of a constant matrix in accordance to the theorem of
section 2.

Finally, the most general line element $(\gamma_{\alpha\beta},
~N,~N^{a})$ satisfying Einstein's equations (2.5), is thus given,
--in matrix notation by:
\begin{displaymath}
\gamma_{most
~general}(t)=\Lambda^{T}(t)\gamma^{Joseph}(h(t))\Lambda(t)
\end{displaymath}
\begin{displaymath}
\Lambda=\left(\begin{array}{ccc}
  \varrho_{1}P(t) & \varrho_{2}P(t) & x(t) \\
  \varrho_{3}P(t) & \varrho_{4}P(t) & y(t) \\
  0 & 0 & 1
\end{array}\right)
\end{displaymath}
where the $\varrho$'s are constant, subject to the condition
$\varrho_{1}\varrho_{4}-\varrho_{2}\varrho_{3}=1$ and:
\begin{displaymath}
  N(t)=\sqrt{|\gamma_{most ~general}|}\dot{h}(t)
\end{displaymath}
\begin{displaymath}
  N^{\alpha}(t)=S^{\alpha}_{\beta}P^{\beta}(h(t))
\end{displaymath}
\begin{displaymath}
  P^{\beta}(h(t))=\{x(t)(\ln\frac{x(t)}{P(t)})^{\cdot}, y(t)(\ln\frac{y(t)}{P(t)})^{\cdot}, (\ln\frac{1}{P(t)})^{\cdot} \}
\end{displaymath}
\begin{displaymath}
  S=\Lambda^{-1}
\end{displaymath}
\begin{displaymath}
\gamma^{Joseph}(h(t))=\left(\begin{array}{ccc}
  \frac{e^{\lambda h(t)}}{q} & 0 & 0 \\
  0 & \frac{e^{-\lambda h(t)}}{q} & 0 \\
  0 & 0 & f
\end{array}\right)
\end{displaymath}
\begin{equation}
\frac{1}{f(h(t))}=q(h(t))=\left\{ \begin{array}{ll}
  \frac{2}{\sqrt{\omega}}\sinh(\pm\sqrt{\omega}h(t)) & \omega>0 \\
  \pm 2h(t) & \omega=0
\end{array}\right.
\end{equation}
\begin{displaymath}
3\omega=\lambda^{2}
\end{displaymath}
where the fourth arbitrary function $h(t)$, accounts for the
time\\
reparametrization covariance.

\section{Discussion}
In this work, we present an approach to the problem of solving
Einstein's equations, for the case of a generic Bianchi-Type
spatially homogeneous spacetime. The approach is not plagued by
the fragmentation characterizing the major part of the existing
rich literature --which is inherited by the diversity of the
various simplifying ansatzen, employed in each case. The key
notion for avoiding this fragmentation, is that of a
Time-Dependent Automorphism Inducing Diffeomorphism; that is, a
general coordinate transformation (2.9), mixing space and time
coordinates, whose action on the line-element of a Bianchi
Geometry, is described by relations (2.14) --viewed as ''gauge''
transformation laws for the dependent variables
$\gamma_{\alpha\beta}(t)$,~$N(t)$ and $N^{\alpha}(t)$. The
investigation for the existence of such G.C.T.'s, leads to the
necessary and sufficient conditions (2.16), (2.17); hence the name
Time-Dependent A.I.D.'s. In each and every Bianchi Type, these
conditions possess a non-empty set of solutions containing
precisely three arbitrary functions of time. A choice of these
arbitrary functions, amounts exactly to a choice for the three
spatial coordinates. Thus, the possibility is offered for
simplifying Einstein equations, --through a simplification of
$\gamma_{\alpha\beta}, N^{\alpha}, N$--, without running the risk
of loss of generality or any sort of incompatibility.\\
Of course, the possible simplifications differ from one Bianchi
Type to another; even within the same Bianchi Type, there are many
possible simplifications --since one, can use the three arbitrary
functions at will. This kinematical freedom, when combined to the
dynamical information --furnished by the linear constraint
equations--, considerably simplifies the form of the line-element
and thus of Einstein's equations, as well. A useful, in our
opinion, irreducible form of the line-element for each Bianchi
Type, is given at the balance of section 2.\\
A statement that applies to all Types is that, using two of the
three arbitrary functions, the scale-factor matrix
$\gamma_{\alpha\beta}(t)$ can always --a priori; i.e. before
solving any classical equations of motion-- be put into a so
called ''symmetric'' \cite{16} form, i.e.
$\gamma_{13}=\gamma_{23}=0$. This applies also for Type II, if we
take instead of the standard form for the structure constants
($C^{1}_{23}=-C^{1}_{32}=1$, all other vanish) the equivalent
version $C^{3}_{12}=-C^{3}_{21}=1$, all other vanish. If this
''symmetric'' form, is then substituted into the linear equations,
and the third arbitrariness is used, considerable restrictions
among $N^{\alpha}$'s and the remaining $\gamma_{\alpha\beta}$'s
are obtained, as presented in detail at the end of section 2.
Furthermore, with the help of the essential arbitrary constants in
$\Lambda^{\alpha}_{\beta}$,
we can diagonalize $\gamma_{\alpha\beta}(t)$, on mass-shell.\\
For all Bianchi Types, the shift vector $N^{\alpha}$, can always
be set to zero --with the help of Time-Dependent A.I.D.'s, and the
linear equations. One could of course, rely on the well-known
existence of Gauss-normal coordinates \cite{2}, and argue that
this should be true. However, in this work, the explicit
realization of this fact is presented; what is more important, is
that the vanishing of $N^{\alpha}$, is accomplished without
spoiling manifest spatial homogeneity. The interplay between
line-elements with and without shift, established through
Time-Dependent A.I.D.'s --see (2.14b)--, raises the need to
reexamine the set of existing solutions -with respect to physical
equivalence, among each other. In particular, many tilded and
untilded fluid solutions \cite{18}, may proven to be G.C.T.
related --and thus physically indistinguishable.

Except of the three arbitrary functions of time, of considerable
importance, are also the (non absorbable in the shift) arbitrary
constants, appearing in the solutions to (2.16) and (2.17). The
number of these constants, varies for different Bianchi Types. The
very interesting fact, is that when this number is subtracted from
the number of constants, given by Peano's theorem, --after the
freedom in arbitrary functions of time, has been fully
exhausted--, the resulting number of the --finally-- remaining
constants, equals, for each and every Bianchi Type, to the number
of expected essential constants --see \cite{19}, p. 211. This,
permits us to conclude that the gauge symmetry transformations
(2.14) --with (2.16) and (2.17), holding-- are, essentially,
unique. It is also noteworthing, that the existence of these
constant parameters, helps to rectify a defect from which, the
previous approach of Jantzen, is suffering; that of an uneven
passage, from the lower to the higher Bianchi Types, owing to the
change of the dimension of the invoked symmetry group \cite{19};
indeed, the arbitrary functions of time are thus varying with
dim[SAut(G)], from 8 (Type I), to 5 (Type II and V), to 3 (higher
Types). This situation, is rather unsatisfactory, since we know
that the independent or dynamical degrees of freedom for the
gravitational field, are 2 --per space point. Thus, in cosmology,
we expect 2 independent functions
of time --irrespective of Bianchi Type.\\
In contrast to this state of affairs, the solutions to (2.16) and
(2.17), contain exactly 3 arbitrary functions of time, which
together with the arbitrary function --owing to the time
reparametrization covariance of equations (2.5)--, leave us with
$6(\gamma_{\alpha\beta})-4=2$ arbitrary functions, in all Bianchi
Types. The required sensitivity, of the method, to the particular
isometry group, is represented by the extra constant parameters
--as explained.

It is in this remarkable way, that General Relativity manages to
encode the memory of spatial G.C.T. covariance, in the set of the
reduced equations (2.5), where only functions of time and their
derivatives appear. This encoding also persists in the actions
--when these actions exist--, and leads to important grouping of
$\gamma_{\alpha\beta}$'s, into the three solutions:
$x^{1}=C^{\alpha}_{\mu\nu}C^{\beta}_{\rho\sigma}\gamma^{\mu\rho}\gamma^{\nu\sigma}\gamma_{\alpha\beta},~
x^{2}=C^{\alpha}_{\beta\delta}C^{\delta}_{\nu\alpha}\gamma^{\beta\nu},~x^{3}=\gamma$
of the quantum linear constraints \cite{13,20}. When a truly
scalar Hamiltonian exists \cite{13,21}, the wavefunction depends
only on the $q^{i}$'s:
\begin{displaymath}
q^{1}=\frac{m^{\alpha\beta}\gamma_{\alpha\beta}}{\sqrt{\gamma}},~~~q^{2}=\frac{(m^{\alpha\beta}\gamma_{\alpha\beta})^{2}}{2\gamma}
-\frac{1}{4}C^{\alpha}_{\mu\nu}C^{\beta}_{\rho\sigma}\gamma^{\mu\rho}\gamma^{\nu\sigma}\gamma_{\alpha\beta}
,~~~q^{3}=\frac{m}{\sqrt{\gamma}}
\end{displaymath}
which completely and irreducibly, determine a spatial
three-geometry.

To summarize, the system (2.5), admits solutions containing in
each and every Bianchi Type, exactly four unspecified functions of
time. One, corresponds to the freedom of changing the time
coordinate; three, correspond to the freedom of changing the
spatial coordinates via Time-Dependent A.I.D.'s. The action of
such a transformation on the line-element, and on the system of
equations (2.5), is described by relations (2.14), (2.18). Thus,
one does not actually need to calculate the simplifying G.C.T.'s;
one simply uses (2.14), simplifies the equations, solves them
completely, and finally inverts the transformation thereby
obtaining the entire space of solutions. It is in this sense,
that the closed form of the line elements presented in section 3,
exhaust the space of classical solutions --for the case of Bianchi
types II and V.

\vspace{1cm} \noindent \textbf{Acknowledgments} We would like to
thank Prof.~M.A.H. MacCallum (private communication with T.C.),
whose severe criticism on an earlier version of our work, has
helped us to clarify the issue of essential constants --in the
context of the present work. He also brought to our attention, the
existing literature concerning the diagonalization of positive
definite symmetric forms, via the Lorentz group.

The earlier version of this work is part of the 1995 PENED program
''Quantum and Classical Gravity - Black Holes'' (no. 512)
supported by the General Secretariat for the Research and
Technology of the Hellenic Department of Industry, Research and
Technology. Two of the authors (G. Kofinas and A. Paschos) were
partially supported by the Hellenic Fellowship Foundation (I.K.Y.)
All authors, acknowledge financial support by the University of
Athens' Special Account for the Research.

\newpage

\emph{The number of arbitrary constants appearing in general
solution for each Bianchi Type --vacuum model--, is given in the
following table --depicted in the first of \cite{19}, pp. 211:}

\begin{center}
\begin{tabular}{|l|c|}
\hline \hline
Bianchi Type & \# of the essential constants \\
\hline
I & 1 \\
II & 2 \\
$VI_{0}$, $VII_{0}$ & 3 \\
VIII, IX & 4 \\
IV & 3 \\
V & 1 \\
$VI_{h}$ ($h\neq$ -1/9) & 3 \\
$VI_{-1/9}$ & 4 \\
$VII_{h}$ & 3 \\
\hline \hline
\end{tabular}
\end{center}

\newpage
\appendix
\section{Appendix}

In \cite{22}, the following Theorem, is given:\\
\textit{''Let two symmetric forms A and B, be given, on a
n-dimensional linear vector space V. If one of them --say A-- is
non singular, then there is a base in V in which both A and B, are
diagonal, if and only if, the mapping $A^{-1}B$, possesses n-real
eigenvalues.''}

Thus, if we take the pair $\gamma_{\alpha\beta}$,
$\eta_{\alpha\beta}$, it suffices to prove that
$\eta^{\alpha\varrho}\gamma_{\varrho\beta}$, has n-real
eigenvalues. In what follows, for the sake of completeness, we
give a proof of the entire statement that a positive definite
matrix $\gamma_{\alpha\beta}$, can be diagonalized via the Lorentz
group.
\underline{\textbf{Theorem}}\\
Let $\gamma$ be a positive definite $n\times n$ real matrix. Then,
there exists a Lorentz matrix $\Lambda$, such that:
\begin{equation}
\Lambda^{T}\gamma\Lambda=\Delta
\end{equation}
where $\Delta$ a diagonal matrix.\\
\emph{Note}: Since $\Lambda^{T}=\eta \Lambda^{-1}\eta$, where
$\eta$ is the Minkowski metric, (A.1) may be written as
\begin{equation}
\Lambda^{-1}\eta\gamma\Lambda=\eta\Delta
\end{equation}
In order to prove (A.2) it is useful to write it equivalently
using the notation employed with linear mappings. To do that, we
consider an $n$-dimensional real linear space $V$ with basis
$(e_{1},~e_{2},\ldots,~e_{n})$. The scalar product in this space
is defined as $<~,~>~:~V\times V \rightarrow  \Re$, with
$<e_{\alpha},~e_{\beta}>=\eta_{\alpha\beta}$. The matrix
$\eta\gamma$ defines a mapping $f:~ V \rightarrow V$ through the
relation:
\begin{displaymath}
f(e_{\alpha})=\sum^{n}_{\beta=1}(\eta\gamma)_{\alpha\beta}e_{\beta}
\end{displaymath}
The following will prove useful later on:\\
1) If $M\subseteq V$ then $V = M \oplus M^{\perp}$ ~\cite{23}.\\
2) A mapping $f:~ V \rightarrow V$ is called self-dual, if\\
$<f(x),~y>=<x,~f(y)>$ for every $x,~y ~\in ~V$. We may prove that
the mapping $f$ defined through the matrix $\eta\gamma$ is
self-dual. Indeed:
\begin{displaymath}
\left.
\begin{array}{c}
  <f(x),~y>=<y,~f(x)>=y^{T}\eta\eta\gamma x=y^{T}\gamma x \\
  <x,~f(y)>=x^{T}\eta\eta\gamma y=x^{T}\gamma y =y^{T} \gamma x
\end{array} \right\}
\end{displaymath}
\begin{displaymath}
\Rightarrow ~<f(x),~y>=<x,~f(x)>
\end{displaymath}
3) If $M\subseteq V$ is an invariant subspace of $V$ with respect
to a self-dual mapping $f$ then $M^{\perp}$ is also an invariant
subspace of $V$. Indeed, let $b \in M^{\perp}$ and $m \in M$.
Since $M$ is an invariant subspace, it follows that:
\begin{displaymath}
f(m) \in M \Rightarrow ~<f(m),~b>=0 \Rightarrow ~<m,~f(b)>=0 ~
\forall ~m \in M
\end{displaymath}
\begin{displaymath}
\Rightarrow f(b) \in M^{\perp}
\end{displaymath}
Equation (A.2) states the fact that there exists an orthonormal
basis of $V$ consisting of the eigenvalues of $f$. If (A.2) holds
then the non-vanishing elements of the real diagonal matrix
$\eta\Delta$ will be eigenvalues of $\eta\gamma$. Thus, we have
to prove that the eigenvalues of $\eta\gamma$ are real. Indeed,
the following theorem holds:\\
\underline{\textbf{Theorem}}\\
If $\gamma$ is a positive definite symmetric matrix, then
$\eta\gamma$ has real eigenvalues.\\
\underline{\textbf{Proof}}\\
Let $\lambda=\alpha+\beta j$, $~\beta\neq 0$ a complex eigenvalue
of $\eta\gamma$ and $u \neq 0$ the corresponding complex right
eigenvector. Since $\eta$ is invertible, there exists a $v=x+yj$,
$~x,~y,~\in \Re^{n}$ such that $u=\eta v$. We have:
\begin{displaymath}
\eta\gamma u=\lambda u \Leftrightarrow \eta\gamma\eta
v=\lambda\eta v \Leftrightarrow
\end{displaymath}
\begin{equation}
\eta\gamma\eta x=\alpha\eta x-\beta\eta y
\end{equation}
\begin{equation}
\eta\gamma\eta y=\alpha\eta y+\beta\eta x
\end{equation}
Equations (A.3), (A.4) imply respectively:
\begin{displaymath}
y^{T}\eta\gamma\eta x=\alpha <y,~x>-\beta <y,~y>
\end{displaymath}
\begin{displaymath}
x^{T}\eta\gamma\eta y=\alpha <x,~y>+\beta <x,~x>
\end{displaymath}
The last two equations have their left-hand sides equal (since
$\eta\gamma\eta$ is symmetric), hence:
\begin{equation}
\beta (<x,~x>+<y,~y>)=0 \Rightarrow <y,~y>=-<x,~x>
\end{equation}
Since $\gamma$ is positive definite, $\eta\gamma\eta$ is positive
definite as well. Then:
\begin{equation}
x^{T}\eta\gamma\eta x ~\geq ~0 \stackrel{(A.3)}{\Rightarrow}
\alpha <x,~x>-\beta <x,~y> ~\geq ~0
\end{equation}
\begin{equation}
\begin{split}
y^{T}\eta\gamma\eta y ~\geq ~0 \stackrel{(A.4)}{\Rightarrow}
&\alpha <y,~y>+\beta <y,~x> ~\geq ~0
\stackrel{(A.5)}{\Rightarrow}\\& \alpha <x,~x>-\beta <x,~y> ~\leq
~0
\end{split}
\end{equation}
>From (A.6), (A.7) we get:
\begin{equation}
\alpha <x,~x>=\beta <x,~y>
\end{equation}
Through (A.5), (A.8), equations (A.3), (A.4) imply:
\begin{displaymath}
\begin{array}{c}
  x^{T}\eta\gamma\eta x =0 \\
  y^{T}\eta\gamma\eta y =0
\end{array}
\end{displaymath}
and, since $\eta\gamma\eta$ is positive definite we conclude that
$x=0$ and $y=0$, i.e. $u=0$, contradicting our initial assumption
$u \neq 0$. Therefore $\beta$ has to vanish and thus we have
proved the reality of $\lambda$.\\
For the eigenvectors of $\eta\gamma$, we can prove that they have
a non-zero norm. Indeed, let $x$ be an eigenvector of
$\eta\gamma$, i.e.
\begin{displaymath}
\eta\gamma x =\lambda x \Rightarrow \gamma x = \lambda\eta x
\Rightarrow x^{T}\gamma x=\lambda x^{T}\eta x =\lambda <x,~x>
\end{displaymath}
Since $\gamma$ is positive definite and $x \neq 0$ we have
$x^{T}\gamma x > 0$, so that $<x,~x> \neq 0$.\\
We are now in position to prove a spectral theorem for a mapping
$f$ with real eigenvalues.\\
\underline{\textbf{Theorem}}\\
If $f: V \rightarrow V$ is a self-dual mapping with real
eigenvalues, then $V$ has an orthonormal basis consisting of the eigenvectors of $f$.\\
\underline{\textbf{Proof}}\\
Let $\lambda$ be an eigenvalue of $f$, u the corresponding
eigenvector and $M=[u]$ the one-dimensional subspace spanned by
$u$. Obviously, $M$ is an invariant subspace of $V$ with respect
to $f$.\\
According to 1), we have $V=M\oplus M^{\perp}$. As implied by 2)
and 3), $M^{\perp}$ is also an invariant subspace and thus $f$
induces a self-dual mapping onto $M^{\perp}$. Hence, we can apply
induction and show that:
\begin{displaymath}
V=M_{1}\oplus M_{2}\oplus \ldots \oplus M_{n}
\end{displaymath}
where the $M_{\alpha}$ are one-dimensional invariant subspaces
orthogonal to each other. Since $u_{\alpha}$ is an eigenvector of
$\eta\gamma$, it holds that $<u,~u> \neq 0$, as proved above. We
can thus promote the orthogonal basis to an orthonormal set
$(\hat{u}_{1},~\hat{u}_{2},\dots,~\hat{u}_{n})$. The
transformation connecting this orthonormal basis to the initial
orthonormal basis $(e_{1},~e_{2},\ldots,~e_{n})$ is the matrix
$\Lambda$ sought for in the first theorem, relations (A.2) and
(A.1).

\end{document}